\documentclass[11pt]{style/orca}
\pdfoutput=1
\usepackage[toc,page]{appendix}
\usepackage{amsmath}
\usepackage{bigints}
\usepackage{caption}
\usepackage{longtable}
\usepackage{subcaption}

\usetikzlibrary{decorations.markings, calc, positioning}
\hypersetup{
    pdfauthor={Simone Hu, Oliver Schnetz, Jim Shaw, Karen Yeats},
    pdftitle={Further investigations into the graph theory of phi4-periods and the c2 invariant},
}
\DeclareSymbolFont{Letters}{OML}{cmm}{m}{it}
\DeclareMathSymbol{\psi}{\mathalpha}{Letters}{"20}

\newcommand{\DTR}{%
  \begin{tikzpicture}[main node/.style={inner sep=0,minimum size =.12cm,circle,fill=black!80,draw},node distance = 0.75cm,scale=0.7]
    \begin{scope}
      \node[main node, label=above:$C$] (a) at (0,0) {};
      \node[main node, label=above right:$A$] (b) at (2,1) {};
      \node[main node, label=below right:$B$] (c) at (2,-1) {};
      \node[main node, label=above:$D$] (d) at (4,0) {};
      \node[main node] (e) at (2,2.5) {};
      \node[main node] (f) at (2,-2.5) {};
      \node[main node] (g) at (-1,1) {};
      \node[main node] (h) at (-1,-1) {};
      \node[main node] (i) at (5,1) {};
      \node[main node] (j) at (5,-1) {};

      \draw (a) -- (b) -- (d) -- (c) -- (a);
      \draw (e) -- (b) -- (c) -- (f);
      \draw (g) -- (a) -- (h);
      \draw (i) -- (d) -- (j);
    \end{scope}
    \node at (7,0) {$\rightsquigarrow$};
    \begin{scope}[shift={(10,0)}]
      \node[main node, label=above:$C$] (a) at (0,0) {};
      \node[main node, label=above right:$A$] (b) at (2,1) {};
      \node[main node, label=above:$D$] (d) at (4,0) {};
      \node[main node] (e) at (2,2.5) {};
      \node[main node] (f) at (2,-2.5) {};
      \node[main node] (g) at (-1,1) {};
      \node[main node] (h) at (-1,-1) {};
      \node[main node] (i) at (5,1) {};
      \node[main node] (j) at (5,-1) {};

      \draw (a) -- (b) -- (d) -- (a);
      \draw (e) -- (b) -- (f);
      \draw (g) -- (a) -- (h);
      \draw (i) -- (d) -- (j);
    \end{scope}
  \end{tikzpicture}
}

\newcommand{\CaseOne}{%
  \begin{tikzpicture}[main node/.style={inner sep=0,minimum size =.12cm,circle,fill=black!80,draw},node distance = 0.75cm,scale=0.7]
    \begin{scope}
      \node[main node, style={fill=white}] (a) at (0,0) {};
      \node[main node, label=above right:$\infty$] (b) at (2,1) {};
      \node[style={inner sep=2.5,circle,draw}] (v) at (2,1) {};
      \node[main node, style={fill=white}] (c) at (2,-1) {};
      \node[main node, style={fill=white}] (d) at (4,0) {};
      \node[main node, style={fill=white}] (e) at (2,2.5) {};
      \node[main node, label=below:$C$] (f) at (2,-2.5) {};
      \node[main node, label=left:$A$] (g) at (-1,1) {};
      \node[main node, label=left:$B$] (h) at (-1,-1) {};
      \node[main node, label=right:$E$] (i) at (5,1) {};
      \node[main node, label=right:$D$] (j) at (5,-1) {};

      \draw (a) -- (b) -- (d) -- (c) -- (a);
      \draw (e) -- (b) -- (c) -- (f);
      \draw (g) -- (a) -- (h);
      \draw (i) -- (d) -- (j);
    \end{scope}
    \node at (7,0) {$\rightsquigarrow$};
    \begin{scope}[shift={(10,0)}]
      \node[main node, style={fill=white}] (a) at (0,0) {};
      \node[main node, label=above right:$\infty$] (b) at (2,1) {};
      \node[style={inner sep=2.5,circle,draw}] (v) at (2,1) {};
      \node[main node, style={fill=white}] (d) at (4,0) {};
      \node[main node, style={fill=white}] (e) at (2,2.5) {};
      \node[main node, label=below:$C$] (f) at (2,-2.5) {};
      \node[main node, label=left:$A$] (g) at (-1,1) {};
      \node[main node, label=left:$B$] (h) at (-1,-1) {};
      \node[main node, label=right:$E$] (i) at (5,1) {};
      \node[main node, label=right:$D$] (j) at (5,-1) {};

      \draw (a) -- (b) -- (d) -- (a);
      \draw (e) -- (b) -- (f);
      \draw (g) -- (a) -- (h);
      \draw (i) -- (d) -- (j);
    \end{scope}
    \node[rotate=270] at (2, -4) {$\rightsquigarrow$};
    \node[rotate=270] at (12, -4) {$\rightsquigarrow$};
    \begin{scope}[shift={(0, -6)}]
      \node[main node, style={fill=white}] (a) at (0,0) {};
      \node[main node, style={fill=white}] (c) at (2,-1) {};
      \node[main node, style={fill=white}] (d) at (4,0) {};
      \node[main node, label=below:$C$] (f) at (2,-2.5) {};
      \node[main node, label=left:$A$] (g) at (-1,1) {};
      \node[main node, label=left:$B$] (h) at (-1,-1) {};
      \node[main node, label=right:$E$] (i) at (5,1) {};
      \node[main node, label=right:$D$] (j) at (5,-1) {};

      \draw (d) -- node[above]{$3$} (c) -- node[above]{$2$} (a);
      \draw (c) -- node[left]{$1$} (f);
      \draw (g) -- node[above]{$6$} (a) -- node[below]{$7$} (h);
      \draw (i) -- node[above]{$4$} (d) -- node[below]{$5$} (j);
    \end{scope}
    \begin{scope}[shift={(10,-6)}]
      \node[main node, style={fill=white}] (a) at (0,0) {};
      \node[main node, style={fill=white}] (d) at (4,0) {};
      \node[main node, label=below:$C$] (f) at (2,-2.5) {};
      \node[main node, label=left:$A$] (g) at (-1,1) {};
      \node[main node, label=left:$B$] (h) at (-1,-1) {};
      \node[main node, label=right:$E$] (i) at (5,1) {};
      \node[main node, label=right:$D$] (j) at (5,-1) {};

      \draw (d) -- node[above]{$1$} (a);
      \draw (g) -- node[above]{$6$} (a) -- node[below]{$7$} (h);
      \draw (i) -- node[above]{$4$} (d) -- node[below]{$5$} (j);
    \end{scope}
  \end{tikzpicture}
}

\newcommand{\CaseOneBlobNone}{%
  \node[main node, label=below:$C$] (f) at (0,0) {};
  \node[main node, label=above:$A$] (g) at (-1,2) {};
  \node[main node, label=right:$B$] (h) at (-1,1) {};
  \node[main node, label=above:$E$] (i) at (1,2) {};
  \node[main node, label=left:$D$] (j) at (1,1) {};

  \draw (g) to [bend right=60] (h) to [bend right=60] (f) to [bend right=60] (j) to [bend right=60] (i);
  \draw (g) to [bend right=130, in=310, looseness=6.5] (i);
}

\newcommand{\CaseOneBlobL}{%
  \node[main node] (f) at (0,0) {};
  \node[main node] (h) at (-1,1.5) {};
  \node[main node] (i) at (1,2) {};
  \node[main node] (j) at (1,1) {};
  \node (v) at (-1.75, 1.5) {};

  \draw (h) to [out=135, in=230, looseness=15] (h);
  \draw (h) to [bend right=75] (f) to [bend right=60] (j) to [bend right=60] (i);
  \draw (h) to [out=110, in=65, looseness=1.5] (v) to [bend right=120, in=310, looseness=4.5] (i);
}

\newcommand{\CaseOneDTBlob}{%
  \begin{tikzpicture}[main node/.style={inner sep=0,minimum size =.12cm,circle,fill=black!80,draw},node distance = 0.75cm,scale=0.7]
    \begin{scope}[shift={(-10,0)}]
      \CaseOneBlobNone
    \end{scope}
    \node at (-7,0.5) {$\times$};
    \node at (-6,0.5) {$\Bigg($};
    \begin{scope}[shift={(3.5,0)}]
      \reflectbox{
        \CaseOneBlobL
        \node[above=1pt of i] {\reflectbox{$A$}};
        \node[left=1pt of j] {\reflectbox{$B$}};
        \node[below=1pt of f] {\reflectbox{$C$}};
        \node[above=4pt of h] {\reflectbox{$D,E$}};
      }
    \end{scope}
    \node at (-0.75,0.5) {$\bigcap$};
    \begin{scope}[shift={(2,0)}]
      \CaseOneBlobL
      \node[above=1pt of i] {$E$};
      \node[left=1pt of j] {$D$};
      \node[below=1pt of f] {$C$};
      \node[above=4pt of h] {$A,B$};
    \end{scope}
    \node at (4.5,0.5) {$\Bigg)$};
  \end{tikzpicture}
}

\newcommand{\CaseTwo}{%
  \begin{tikzpicture}[main node/.style={inner sep=0,minimum size =.12cm,circle,fill=black!80,draw},node distance = 0.75cm,scale=0.7]
    \begin{scope}
      \node[main node] (a) at (0,0) {};
      \node[style={inner sep=2.5,circle,draw}, label=above:$\infty$] (v) at (0,0) {};
      \node[main node, style={fill=white}] (b) at (2,1) {};
      \node[main node, style={fill=white}] (c) at (2,-1) {};
      \node[main node, label=above:$A$] (d) at (4,0) {};
      \node[main node, label=above:$B$] (e) at (2,2.5) {};
      \node[main node, label=below:$C$] (f) at (2,-2.5) {};
      \node[main node] (g) at (-1,1) {};
      \node[main node] (h) at (-1,-1) {};
      \node[main node] (i) at (5,1) {};
      \node[main node] (j) at (5,-1) {};

      \draw (a) -- (b) -- (d) -- (c) -- (a);
      \draw (e) -- (b) -- (c) -- (f);
      \draw (g) -- (a) -- (h);
      \draw (i) -- (d) -- (j);
    \end{scope}
    \node at (7,0) {$\rightsquigarrow$};
    \begin{scope}[shift={(10,0)}]
      \node[main node] (a) at (0,0) {};
      \node[style={inner sep=2.5,circle,draw}, label=above:$\infty$] (v) at (0,0) {};
      \node[main node, style={fill=white}] (b) at (2,1) {};
      \node[main node, label=above:$A$] (d) at (4,0) {};
      \node[main node, label=above:$B$] (e) at (2,2.5) {};
      \node[main node, label=below:$C$] (f) at (2,-2.5) {};
      \node[main node] (g) at (-1,1) {};
      \node[main node] (h) at (-1,-1) {};
      \node[main node] (i) at (5,1) {};
      \node[main node] (j) at (5,-1) {};

      \draw (a) -- (b) -- (d) -- (a);
      \draw (e) -- (b) -- (f);
      \draw (g) -- (a) -- (h);
      \draw (i) -- (d) -- (j);
    \end{scope}
    \node[rotate=270] at (2, -4) {$\rightsquigarrow$};
    \node[rotate=270] at (12, -4) {$\rightsquigarrow$};
    \begin{scope}[shift={(0,-9)}]
      \node[main node, style={fill=white}] (b) at (1,1) {};
      \node[main node, style={fill=white}] (c) at (3,1) {};
      \node[main node, label=above:$A$] (d) at (2,2.5) {};
      \node[main node, label=below:$B$] (e) at (0,0) {};
      \node[main node, label=below:$C$] (f) at (4,0) {};
      \node[main node] (i) at (1,3.5) {};
      \node[main node] (j) at (3,3.5) {};

      \draw (b) -- node[left]{$1$} (d) -- node[right]{$3$} (c);
      \draw (e) -- node[left]{$4$} (b) -- node[below]{$2$}(c) -- node[right]{$5$}(f);
      \draw (i) -- (d) -- (j);
    \end{scope}
    \begin{scope}[shift={(11,-9)}]
      \node[main node, style={fill=white}] (b) at (1,1) {};
      \node[main node, label=above:$A$] (d) at (1,3) {};
      \node[main node, label=below:$B$] (e) at (0,0) {};
      \node[main node, label=below:$C$] (f) at (2,0) {};
      \node[main node] (i) at (0,4) {};
      \node[main node] (j) at (2,4) {};

      \draw (b) -- node[left]{$1$} (d);
      \draw (e) -- node[above]{$4$} (b) -- node[above]{$5$} (f);
      \draw (i) -- (d) -- (j);
    \end{scope}
  \end{tikzpicture}
}

\newcommand{\CaseTwoBlobNone}{%
  \node[main node, label=above:$A$] (f) at (0,0) {};
  \node[main node] (g) at (-1,1) {};
  \node[main node, label=above:$B$] (h) at (-1,-1) {};
  \node[main node] (i) at (1,1) {};
  \node[main node, label=above:$C$] (j) at (1,-1) {};

  \draw (g) -- (f) -- (i);
  \draw (g) to [bend right=60] (h) to [bend right=60] (j) to [bend right=60] (i);
  \draw (g) to [bend right=135, in=315, looseness=7] (i);
}

\newcommand{\CaseTwoBlobL}{%
  \node[main node] (f) at (0,0) {};
  \node[main node] (g) at (-1,1) {};
  \node[main node] (i) at (1,1) {};
  \node[main node] (j) at (1,-1) {};

  \draw (g) to (f) to (i);
  \draw (g) to [bend right=75, looseness=2] (f) to [bend right=75, looseness=2] (j) to [bend right=60] (i);
  \draw (g) to [bend right=135, in=315, looseness=7] (i);
}

\newcommand{\CaseTwoBlob}{%
  \begin{tikzpicture}[main node/.style={inner sep=0,minimum size =.12cm,circle,fill=black!80,draw},node distance = 0.75cm,scale=0.7]
    \begin{scope}[shift={(-10,0)}]
      \CaseTwoBlobNone
    \end{scope}
    \node (x) at (-7,-0.5) {$\times$};
    \node at (-6,-0.5) {$\Bigg($};
    \begin{scope}[shift={(3.5,0)}]
      \reflectbox{
        \CaseTwoBlobL
        \node[above=4pt of f] {\reflectbox{$A,C$}};
        \node[above=1pt of j] {\reflectbox{$B$}};
      }
    \end{scope}
    \node at (-0.75,-0.5) {$\bigcap$};
    \begin{scope}[shift={(2,0)}]
      \CaseTwoBlobL
      \node[above=4pt of f] {$A,B$};
      \node[above=1pt of j] {$C$};
    \end{scope}
    \node at (4.5,-0.5) {$\Bigg)$};
  \end{tikzpicture}
}

\newcommand{\GrphOne}{%
  \begin{tikzpicture}[main node/.style={inner sep=0,minimum size =.12cm,circle,fill=black!80,draw},node distance = 0.75cm,scale=0.4]
    \node[main node] (a) at (0,0) {};
    \node[main node] (b) at (6,0) {};
    \node [above left=5mm of a] (a1) {};
    \node [below left=5mm of a] (a2) {};
    \node [above right=5mm of b] (b1) {};
    \node [below right=5mm of b] (b2) {};

    \draw (a) -- (a1);
    \draw (a) -- (a2);
    \draw (b) -- (b1);
    \draw (b) -- (b2);
    \draw (a) to [bend left] node[above]{$\alpha_1$} (b);
    \draw (b) to [bend left] node[below]{$\alpha_2$} (a);
  \end{tikzpicture}
}
\newcommand{\CmplOne}{%
  \begin{tikzpicture}[main node/.style={inner sep=0,minimum size =.12cm,circle,fill=black!80,draw},node distance = 0.75cm,scale=0.4]
    \node[main node] (a) at (0,0) {};
    \node[main node] (b) at (6,0) {};
    \node[main node, label=above:$v$] (c) at (3,4) {};

    \draw (a) to [bend left=30] (b);
    \draw (b) to [bend left=30] (a);
    \draw (a) to [bend left=30] (c);
    \draw (c) to [bend left=30] (a);
    \draw (b) to [bend left=30] (c);
    \draw (c) to [bend left=30] (b);
  \end{tikzpicture}
}

\newcommand{\GrphTwoB}{%
  \begin{tikzpicture}[main node/.style={inner sep=0,minimum size =.12cm,circle,fill=black!80,draw},node distance = 0.75cm,scale=0.4]
    \node[main node] (a) at (0,3) {};
    \node[main node] (b) at (5,3) {};
    \node[main node] (c) at (10,3) {};
    \node [above left=5mm of a] (a1) {};
    \node [below left=5mm of a] (a2) {};
    \node [above right=5mm of c] (c1) {};
    \node [below right=5mm of c] (c2) {};

    \draw (a) -- (a1);
    \draw (a) -- (a2);
    \draw (c) -- (c1);
    \draw (c) -- (c2);
    \draw (a) to [bend left] (b);
    \draw (b) to [bend left] (a);
    \draw (c) to [bend left] (b);
    \draw (b) to [bend left] (c);
  \end{tikzpicture}
}

\newcommand{\CmplTwoB}{%
  \begin{tikzpicture}[main node/.style={inner sep=0,minimum size =.12cm,circle,fill=black!80,draw},node distance = 0.75cm,scale=0.4]
    \node[main node] (a) at (0,0) {};
    \node[main node] (b) at (6,0) {};
    \node[main node] (c) at (6,6) {};
    \node[main node] (d) at (0,6) {};
    \node (l1) at (3,8) {};
    \node (l2) at (3,-2) {};

    \draw (a) to [bend left] (b);
    \draw (b) to [bend left] (a);
    \draw (c) to [bend left] (d);
    \draw (d) to [bend left] (c);
    \draw (a) to [bend left] (d);
    \draw (d) to [bend left] (a);
    \draw (b) to [bend left] (c);
    \draw (c) to [bend left] (b);
    \draw[dashed] (l1) -- (l2);
  \end{tikzpicture}
}
\newcommand{\FourierTwist}{%
  \begin{tikzpicture}[main node/.style={inner sep=0,minimum size =.12cm,circle,fill=black!80,draw},dual node/.style={inner sep=0,minimum size =.12cm,circle,fill=white!80,draw},node distance = 0.75cm,scale=0.4]
    \begin{scope}[shift={(9,0)}]
      \node[main node, label=below:$\infty$] (a) at (0,0) {};
      \node[main node] (b) at (3.75,2.5) {};
      \node[main node] (c) at (3.75,5) {};
      \node[main node, label=above:$0$] (d) at (0,7.5) {};
      \node[main node] (e) at (-3.75,6.25) {};
      \node[main node] (f) at (-5,3.75) {};
      \node[main node] (g) at (-3.75,1.25) {};
      \node[main node, label=above:$1$] (h) at (0,5) {};
      \node[main node, label=above:$z$] (i) at (0,2.5) {};

      \draw (a) -- (b) -- (c) -- (d) -- (e) -- (f) -- (g) -- (a);
      \draw (a) -- (c) -- (h) -- (b) -- (i) -- (g) -- (h) -- (e) -- (a);
      \draw (d) -- (f) -- (i);
      \draw (d) to [bend left] (i);
    \end{scope}
    \node at (16,3.5) {$\rightsquigarrow$};
    \node[rotate=225] at (7, -2.5) {$\pmb{\rightsquigarrow}$};
    \begin{scope}[shift={(0,-13)}]
      \node[main node, label=above:$0$] (d) at (-1,7.5) {};
      \node[main node] (e) at (-4.75,6.25) {};
      \node[main node] (f) at (-6,3.75) {};
      \node[main node] (g) at (-4.75,1.25) {};
      \node[main node, label=above:$1$] (h) at (-1,5) {};
      \node[main node, label=above:$z$] (i) at (-1,2.5) {};

      \node[main node, label=above:$0$] (d') at (2,7.5) {};
      \node[main node] (b') at (5.75,2.5) {};
      \node[main node] (c') at (5.75,5) {};
      \node[main node, label=above:$1$] (h') at (2,5) {};
      \node[main node, label=above:$z$] (i') at (2,2.5) {};

      \node[dual node] at (3,6) {};
      \node[dual node] at (3,3.5) {};
      \node[dual node] at (4.75,4) {};
      \node[dual node] at (7,4) {};
      \node[dual node] at (3,0) {};
      \node[dual node] at (0.75,5) {};

      \draw (d) -- (e) -- (f) -- (g);
      \draw (i) -- (g) -- (h) -- (e);
      \draw (d) -- (f) -- (i);

      \draw (d') to [bend left=135, out=115, looseness=4.5] (i');
      \draw (b') -- (c') -- (d');
      \draw (c') -- (h') -- (b') -- (i');
      \draw[dashed] (d') to [bend right=60] (h') to [bend right=60] (i');
      \draw[dashed] (d') to [bend right=75, looseness=1.2] (i');
    \end{scope}
    \node at (9,-9.5) {$\rightsquigarrow$};
    \begin{scope}[shift={(15.5,-13)}]
      \node[main node, label=above:$0$] (d) at (0,7.5) {};
      \node[main node] (e) at (-3.75,6.25) {};
      \node[main node] (f) at (-5,3.75) {};
      \node[main node] (g) at (-3.75,1.25) {};
      \node[main node, label=above:$1$] (h) at (0,5) {};
      \node[main node, label=above:$z$] (i) at (0,2.5) {};

      \node[dual node, label=above:$z$] (d') at (3,6) {};
      \node[dual node, label=above:$0$] (h') at (3,3) {};
      \node[dual node, label=above:$1$] (i') at (3,0) {};
      \node[dual node, label=above:$v$] (o') at (1.25,4.5) {};
      \node[dual node] (f') at (4.75,4.5) {};
      \node[dual node] (g') at (7,4.5) {};

      \draw (d) -- (e) -- (f) -- (g);
      \draw (i) -- (g) -- (h) -- (e);
      \draw (d) -- (f) -- (i);

      \draw (i') -- (g') -- (d') -- (f')  -- (g') -- (h') --  (f');
      \draw[dashed] (d') to (o') to (h');
      \draw[dashed] (o') to (i');
      \draw[dotted] (d) -- (h');
      \draw[dotted] (h) -- (i');
      \draw[dotted] (i) -- (d');
    \end{scope}
    \node at (24,-9.5) {$\rightsquigarrow$};
    \begin{scope}[shift={(31.5,-13)}]
      \node[dual node] (b) at (3.75,2.5) {};
      \node[dual node] (c) at (3.75,5) {};
      \node[main node, label=above:$0$] (d) at (0,7.5) {};
      \node[main node] (e) at (-3.75,6.25) {};
      \node[main node] (f) at (-5,3.75) {};
      \node[main node] (g) at (-3.75,1.25) {};
      \node[main node, label=above:$1$] (h) at (0,5) {};
      \node[main node, label=above:$z$] (i) at (0,2.5) {};

      \draw (d) -- (e) -- (f) -- (g);
      \draw (i) -- (g) -- (h) -- (e);
      \draw (d) -- (f) -- (i);
      \draw (i) -- (c) -- (d) -- (b) -- (i);
      \draw (b) -- (c) -- (h);
    \end{scope}
    \node[rotate=135] at (26,-2.5) {$\pmb{\rightsquigarrow}$};
    \begin{scope}[shift={(24,0)}]
      \node[main node, label=below:$\infty$] (a) at (0,0) {};
      \node[main node] (b) at (3.75,2.5) {};
      \node[main node] (c) at (3.75,5) {};
      \node[main node, label=above:$0$] (d) at (0,7.5) {};
      \node[main node] (e) at (-3.75,6.25) {};
      \node[main node] (f) at (-5,3.75) {};
      \node[main node] (g) at (-3.75,1.25) {};
      \node[main node, label=above:$1$] (h) at (0,5) {};
      \node[main node, label=above:$z$] (i) at (0,2.5) {};

      \draw (a) -- (b) -- (c) -- (d) -- (e) -- (f) -- (g) -- (a);
      \draw (d) -- (b) -- (i) -- (c) -- (h) -- (g) -- (i) -- (f) -- (d);
      \draw (a) -- (e) -- (h);
      \draw (h) to [bend left] (a);
    \end{scope}
  \end{tikzpicture}
}

\newcommand{\ConfigOne}{%
  \begin{tikzpicture}[main node/.style={inner sep=0,minimum size =.12cm,circle,fill=black!80,draw},node distance = 0.75cm,scale=0.7]
    \node[main node] (a) at (0,6) {};
    \node[main node] (b) at (0,4) {};
    \node[main node] (c) at (0,2) {};
    \node[main node] (d) at (0,0) {};
    \node[style={inner sep=2.5,circle,draw}, label=above:$\infty$] at (0,0) {};

    \node[above left=2.5mm and 5mm of a] (a1) {};
    \node[below left=2.5mm and 5mm of a] (a2) {};
    \node[above left=2.5mm and 5mm of b] (b1) {};
    \node[below left=2.5mm and 5mm of b] (b2) {};
    \node[above left=2.5mm and 5mm of c] (c1) {};
    \node[below left=2.5mm and 5mm of c] (c2) {};
    \node[above left=2.5mm and 5mm of d] (d1) {};
    \node[below left=2.5mm and 5mm of d] (d2) {};

    \node[above right=2.5mm and 5mm of a] (a3) {};
    \node[below right=2.5mm and 5mm of a] (a4) {};
    \node[above right=2.5mm and 5mm of b] (b3) {};
    \node[below right=2.5mm and 5mm of b] (b4) {};
    \node[above right=2.5mm and 5mm of c] (c3) {};
    \node[below right=2.5mm and 5mm of c] (c4) {};
    \node[above right=2.5mm and 5mm of d] (d3) {};
    \node[below right=2.5mm and 5mm of d] (d4) {};

    \draw (a1) -- (a) -- (a2);
    \draw (a3) -- (a) -- (a4);
    \draw (b1) -- (b) -- (b2);
    \draw (b3) -- (b) -- (b4);
    \draw (c1) -- (c) -- (c2);
    \draw (c3) -- (c) -- (c4);
    \draw (d1) -- (d) -- (d2);
    \draw (d3) -- (d) -- (d4);
  \end{tikzpicture}
}
\newcommand{\ConfigTwo}{%
  \begin{tikzpicture}[main node/.style={inner sep=0,minimum size =.12cm,circle,fill=black!80,draw},node distance = 0.75cm,scale=0.7]
    \node[main node] (a) at (0,6) {};
    \node[style={inner sep=2.5,circle,draw}, label=above:$\infty$] at (0,6) {};
    \node[main node] (b) at (0,4) {};
    \node[main node] (c) at (0,2) {};
    \node[main node] (d) at (0,0) {};

    \node[left=5mm of a] (a1) {};
    \node[left=5mm of b] (b1) {};
    \node[above left=2.5mm and 5mm of c] (c1) {};
    \node[below left=2.5mm and 5mm of c] (c2) {};
    \node[above left=2.5mm and 5mm of d] (d1) {};
    \node[below left=2.5mm and 5mm of d] (d2) {};

    \node[right=5mm of a] (a2) {};
    \node[above right=5mm of a] (a3) {};
    \node[below right=5mm of a] (a4) {};
    \node[right=5mm of b] (b2) {};
    \node[above right=5mm of b] (b3) {};
    \node[below right=5mm of b] (b4) {};
    \node[above right=2.5mm and 5mm of c] (c3) {};
    \node[below right=2.5mm and 5mm of c] (c4) {};
    \node[above right=2.5mm and 5mm of d] (d3) {};
    \node[below right=2.5mm and 5mm of d] (d4) {};

    \draw (a1) -- (a) -- (a2);
    \draw (a3) -- (a) -- (a4);
    \draw (b1) -- (b) -- (b2);
    \draw (b3) -- (b) -- (b4);
    \draw (c1) -- (c) -- (c2);
    \draw (c3) -- (c) -- (c4);
    \draw (d1) -- (d) -- (d2);
    \draw (d3) -- (d) -- (d4);
  \end{tikzpicture}
}
\newcommand{\ConfigThree}{%
  \begin{tikzpicture}[main node/.style={inner sep=0,minimum size =.12cm,circle,fill=black!80,draw},node distance = 0.75cm,scale=0.7]
    \node[main node] (a) at (0,6) {};
    \node[main node] (b) at (0,4) {};
    \node[main node] (c) at (0,2) {};
    \node[main node] (d) at (0,0) {};
    \node[style={inner sep=2.5,circle,draw}, label=above:$\infty$] at (0,0) {};

    \node[left=5mm of a] (a1) {};
    \node[above left=5mm of b] (b1) {};
    \node[left=5mm of b] (b2) {};
    \node[below left=5mm of b] (b3) {};
    \node[above left=2.5mm and 5mm of c] (c1) {};
    \node[below left=2.5mm and 5mm of c] (c2) {};
    \node[above left=2.5mm and 5mm of d] (d1) {};
    \node[below left=2.5mm and 5mm of d] (d2) {};

    \node[right=5mm of a] (a2) {};
    \node[above right=5mm of a] (a3) {};
    \node[below right=5mm of a] (a4) {};
    \node[right=5mm of b] (b4) {};
    \node[above right=2.5mm and 5mm of c] (c3) {};
    \node[below right=2.5mm and 5mm of c] (c4) {};
    \node[above right=2.5mm and 5mm of d] (d3) {};
    \node[below right=2.5mm and 5mm of d] (d4) {};

    \draw (a1) -- (a) -- (a2);
    \draw (a3) -- (a) -- (a4);
    \draw (b1) -- (b) -- (b2);
    \draw (b3) -- (b) -- (b4);
    \draw (c1) -- (c) -- (c2);
    \draw (c3) -- (c) -- (c4);
    \draw (d1) -- (d) -- (d2);
    \draw (d3) -- (d) -- (d4);
  \end{tikzpicture}
}

\bibliography{main}

\title[Further investigations into the graph theory of $\phi^4$-periods and the $c_2$ invariant]{Further investigations into the graph theory of $\phi^4$-periods \\ and the $c_2$ invariant}
\date{\today}
\author{Simone Hu}
\address{Department of Combinatorics and Optimization \newline
  \indent Faculty of Mathematics, University of Waterloo \newline
  \indent Waterloo, ON, Canada, N2L 3G1}
\email{ss2hu@edu.uwaterloo.ca}

\author{Oliver Schnetz}
\address{Department Mathematik \newline
  \indent Emmy-Noether-Zentrum, FAU Erlangen-Nürnberg \newline
  \indent Cauerstr. 11, 91058 Erlangen, Germany} 
\email{schnetz@mi.uni-erlangen.de}

\author{Jim Shaw}
\address{Department of Physics and Astronomy \newline
  \indent Faculty of Applied Science, University of British Columbia \newline
  \indent Vancouver, BC, Canada, V6T 1Z1}
\email{jimshaw@alumni.ubc.ca}

\author{Karen Yeats}
\address{Department of Combinatorics and Optimization \newline
  \indent Faculty of Mathematics, University of Waterloo \newline
  \indent Waterloo, ON, Canada, N2L 3G1}
\email{kayeats@uwaterloo.ca}

\begin{document}
\begin{abstract}
  A Feynman period is a particular residue of a scalar Feynman integral which is both physically and number theoretically interesting.  Two ways in which the graph theory of the underlying Feynman graph can illuminate the Feynman period are via graph operations which are period invariant and other graph quantities which predict aspects of the Feynman period, one notable example is known as the $c_2$ invariant.  We give results and computations in both these directions, proving a new period identity and computing its consequences up to 11 loops in $\phi^4$-theory, proving a $c_2$ invariant identity, and giving the results of a computational investigation of $c_2$ invariants at 11 loops.
\end{abstract}

\maketitle
\setcounter{tocdepth}{1}
\tableofcontents

\setcounter{section}{-1}
\section{Introduction}\label{S:intro}
In perturbative quantum field theory one studies physical processes by expanding in small parameters.  One of the most famous, and still very useful, type of such expansions are expansions indexed by Feynman diagrams. Feynman diagrams are graphs which symbolize particle interactions and each one indexes an integral: its Feynman integral.  For further details see a quantum field theory textbook such as ~\cite{iz}.  Feynman integrals are interesting from many perspectives; physically they are a tool to calculate amplitudes, analytically they are a rich family of very difficult integrals, number theoretically they can often (at least for small graphs) be expressed in terms of multiple zeta values and other arithmetically interesting presumably transcendental numbers, see for instance ~\cite{motives, periods, BrSinform, numfunct} and the references therein.  Feynman integrals and Feynman diagrams also lead to interesting graph theoretic questions.  Sometimes Feynman diagrams motivate new purely combinatorial techniques ~\cite{LZgraphs}.  Other times natural questions on the physics side can be answered by combinatorics, often with questions and results of pure combinatorial interest along the way.  We will be working in this latter direction. 

We will restrict our attention to Feynman integrals of 4-point Feynman diagrams in massless euclidean $\phi^4$-theory in 4-dimensions.  In fact we will further restrict to a particular residue of this integral known as the Feynman period ~\cite{periods, census}.  The Feynman period is essentially the coefficient of the divergence and so for subdivergence free Feynman diagrams the period captures an important renormalization scheme independent part of the Feynman integral.  The number theoretic content of Feynman periods also remains interesting, see for example ~\cite{k3}. 

{}From a graph theoretical side this means that we will be working with graphs which are 4-regular but with the possibility of external edges.  External edges are best thought of as half-edges which add to the degree of their one incident vertex as usual but do not connect to another vertex.  The external edges represent the particles entering or exiting the system.  The period, then, can be defined as an integral directly from the graph.  It is defined and discussed in the next section.  Notably the integral is controlled by the Kirchhoff polynomial of the graph which is a multivariate polynomial given as a sum of spanning trees.  Consequently the Feynman period has an algebro-geometric feel as the variety defined by the vanishing of the Kirchhoff polynomial is central, and a combinatorial feel through the manipulation of spanning trees. 

After setting up the objects and definitions we need, we proceed to show that the graph transformation of taking a planar dual on one side of a small separation in the graph is a period invariant.  This proves many new identities of Feynman periods and we collect the new identities up to 11 loops\footnote{An $\ell$-loop graph is a graph where the dimension of the cycle space is $\ell$.  Another way to say this is that the first Betti number of the graph is $\ell$.} in the first appendix. 

The remainder of the paper considers the $c_2$ invariant, an arithmetic graph invariant defined by one of us in ~\cite{fq}. For a given graph the $c_2$ invariant is a sequence indexed by prime powers (the definition is in the next section). Some very interesting sequences show up including Fourier expansions of modular forms ~\cite{k3, modular, logan}.  Although the Feynman period and the $c_2$ invariant look at the geometry of the Kirchhoff variety from different directions they are closely linked.

There are three directions we could hope to take with the $c_2$ invariant; we could understand its symmetries and properties, we could more precisely understand its connection with the Feynman period, and we could work to compute it.  In the first direction, we look at a known property of the $c_2$ invariant, its invariance under double triangle reductions ~\cite{forest} and tidy up how it relates to a conjectured symmetry known as completion. 

We do not address the second direction, here. We refer the reader to ~\cite{BrownDorync2}.

Regarding the third direction, one of us with Brown in ~\cite{modular} reported on exhaustive calculations for all 4-point $\phi^4$ graphs up to loop order 10 and for small primes.  This was done by using denominator reduction ~\cite{periods} to reduce the number of variables in the polynomial along with further tricks to make the computation tractable, ultimately finishing by directly counting points on a now small polynomial.  However, 11 loops remained out of reach.  Another of us alone and with Chorney ~\cite{yeats2016few, CYgrid, yeats2018study} has used a different approach only for very small primes though applicable to the entirety of certain special families of graphs.  Part of this approach can also be applied to individual graphs and is more tractable for large graphs than the previous approach, though larger primes are less accessible as the complexity growth in the size of the prime is worse.  We use this technique to calculate all 11 loop 4-point $c_2$ invariants up to $p=7$ and many to $p=13$. 

\bpointn{Acknowledgements}%
The authors would like to thank Iain Crump and Erik Panzer for their up-to-date lists of period equivalences and Hepp bounds of 4-point $\phi^4$ graphs up to loop order 11.

KY is supported by an NSERC discovery grant and a Humboldt Fellowship; JS was supported by an NSERC USRA; OS is supported by DFG grant SCHN 1240.

\section{Background}\label{S:background}
In this paper, we will only be considering 4-point Feynman integrals in four-dimensional ($D=4$) massless euclidean $\phi^4$-theory.  Combinatorially, these correspond to graphs with 4 external half-edges and every vertex having degree 4 (where external edges contribute to the degree). 

\bpoint{4-point graphs}%
Let $G$ be a 4-point graph in $\phi^4$-theory, that is a 4-regular graph with 4 external half-edges.  Let $m = \#V$ be the number of vertices in $G$, $n = \#E$ the number of internal edges and $\ell$ the loop order. 

The superficial degree of divergence of a Feynman integral is a measure of how badly the integral diverges as the energies get large.  It is obtained by tallying how many powers of the integration variables are contributed by each edge and vertex, compared to how many integration variables there are.  Consequently, it can be distilled into a purely combinatorial invariant of the graph, see Section 5.2 of ~\cite{ybrief} for a description in a similar language to here. 

In our case, as $\phi^4$-theory is renormalizable in $D = 4$, the superficial degree of divergence of $G$ is
\[ \text{sdd}(G) = 4\ell - 2n = 4 - q \]
where $q$ is the number of external half-edges.  Here we use that there are $2n + q = 4m$ half-edges and Euler's formula for connected graphs,
\[ m - n + \ell = 1, \]
which gives
\[ 4\ell - 2n = 4(1 + n - m) - 2n = 4 + 2n - 4m = 4 - q. \]

As $q=4$, every 4-point graph is logarithmically divergent (that is $\text{sdd}(G) = 0$) and furthermore we get the equality $n = 2\ell$.  Using Euler's formula again gives us that $m = \ell + 1$. 

\bpoint{Periods}\label{SS:periods}%
The standard approach to perturbative quantum field theory begins with a Lagrangian density, which in our case would be
\[ \mathcal{L} = \frac{1}{2}(\partial \phi)^2 - \frac{\lambda}{4!}\phi^4, \]
and then builds the path integral
\[ A = \int D\phi \exp\left(i\int \mathrm{d}^4 x \mathcal{L} + J\phi \right). \]

Expanding in $J$ and taking the coefficient of $J^4$ gives the 4-point function and this itself can be expanded as a series in $\lambda$. Wick's theorem says that this expansion can be calculated by summing over graphs of the type we are working with and each graph contributes its Feynman integral. From this point on we can ignore the standard derivation (and the foundational issues involving the path integral), and simply define the 4-point amplitude as the formal sum of these Feynman integrals viewed as formal integral expressions. Standard results, known in both quantum field theory and enumerative combinatorics ~\cite{JKMlegendre}, allow us to reduce to one-particle irreducible (1PI) graphs\footnote{These are 2-edge-connected graphs in the language of graph theory.} by taking a logarithm and a Legendre transform. Consider then the individual Feynman integrals. These need to be renormalized, but that is not to the point for the present paper, as instead we will simplify matters by restricting to primitive graphs (graphs without subdivergences, see Definition~\ref{def:primitive}). Primitive graphs have finite residues which do not depend on any kinematical parameters and give a renormalization scheme independent contribution to the beta function of the theory.  This residue has come to be known as the \textbf{period} ~\cite{periods, census} of the primitive graph, and is defined below. 

Given a Feynman graph $G$, the Feynman rules tell us how to translate $G$ into its period in different representations, known as the different spaces: position, momentum, parametric and dual parametric.  The first two of these correspond to assigning 4-dimensional vectors to vertices and cycles respectively, with each edge contributing a factor to the integrand; its propagator appropriate to the space. 
Parametric and dual parametric space are slightly different. Now each edge is assigned a variable, completing the trio of variable assignments to vertices, cycles, and edges, but these edge variables are real scalars and are collected into one polynomial rather than each contributing a factor.

To use the Feynman rules, first we arbitrarily orient the edges and cycles of $G$ and suppose $\ell$ is the loop order.
Here we will use the notation $x^2$ to denote the norm squared\footnote{here we conveniently use Euclidean signature, see e.g.\ ~\cite{iz}} $\norm{x}^2$ and $\mathbb{1}$ will represent some fixed choice of unit vector. 

In position space, the variable $x_i$ is attached to vertex $i$.
Each edge $(i,j)$ then gets the propagator $\frac{1}{(x_i - x_j)^2}$.
By setting one vertex to 0, say $x_0$, and one vertex to $\mathbb{1}$, say $x_1$, the period of G is
\[ P_G = \pi^{-2(\ell-1)} \bigintsss \mathrm{d}^4 x_2 \cdots \mathrm{d}^4 x_{\ell} \,\frac{1}{\displaystyle\prod_{e=(i,j)} \left. (x_i - x_j)^2 \right\rvert_{x_0=0, x_1=\mathbb{1}}}.\]
The freedom to set one variable to $0$ and one to $\mathbb{1}$ comes from the fact that the whole integral is invariant under affine linear transformations, so we can translate it to the origin and move it into a standard position (and scale) there. 

In momentum space, now variables $p_i$ are associated with each cycle in an oriented cycle basis of the graph (there are $\ell$ such cycles).
Each edge gets the propagator $\frac{1}{p_e^2}$ where $p_e = \sum \pm p_i$ is the signed sum of the cycles that run through edge $e$, with signs depending on which direction the cycles go through the edge.
By setting one momentum vector (a cycle in the basis) to $\mathbb{1}$, say $p_1$, the period of G is
\[ P_G = \pi^{-2(\ell-1)} \bigintsss \mathrm{d}^4 p_2 \cdots \mathrm{d}^4 p_{\ell} \,\frac{1}{\displaystyle\prod_{e} \left. p_e^2 \right\rvert_{p_1 = \mathbb{1}}}.\]
Here the freedom to set one variable to $\mathbb{1}$ comes from the fact that we can always normalize the momentum variables with respect to one such variable and we have rotational invariance. 

Note that we can transform between position space and momentum space through a Fourier transform.
For both spaces, the domain we are integrating over is all of $\mathbb{R}^4$.

Now, using the Schwinger trick and setting one edge variable to the scalar 1, say $\alpha_1$, we can transform to parametric space to get
\begin{equation}\label{parametric_period}
  P_G = \bigintsss_{\; 0}^{\infty} \mathrm{d}\alpha_2 \cdots \mathrm{d}\alpha_{2\ell} \,\frac{1}{\left. \Psi_G^2 \right\rvert_{\alpha_1=1}}
\end{equation}
where $\alpha_e$ is a variable attached to each edge $e$ in $G$ and
\[ \Psi_G = \sum\limits_{\substack{T \\ \text{spanning tree}}} \prod_{e \notin T} \alpha_e \]
is the \textbf{graph polynomial} or \textbf{Kirchhoff polynomial} of G. 

Finally, dual parametric space is very similar to parametric space, with the only difference being we now take the edges $e \in T$ in the graph polynomial. The graph polynomials for parametric space and dual parametric space are related by a Cremona transformation.

\tpoint{Example} (Triangle graph polynomial)

Let $G$ be a triangle with edges labelled $\alpha_1, \alpha_2$ and  $\alpha_3$. Then the Kirchhoff polynomial of $G$ is $\Psi_G = \alpha_1 + \alpha_2 +\alpha_3$ since there are 3 spanning trees of $G$, each corresponding to the removal of an edge in the triangle. Hence there are three monomials and each monomial corresponds to the edge that was cut to form the spanning tree of the triangle. \\

As all these integrals for $P_G$ relate through some transformation of variables, we must have that these are all equivalent definitions for the same number (if it exists).  Note, in each case, the choice of cycle, vertices or edge to set is arbitrary. See ~\cite{census} for details and proofs.

It turns out that when $G$ is primitive and logarithmically divergent, that is a 4-point graph with no 1PI divergent subgraphs (i.e.\ primitive for the co-product of the renormalization Hopf algebra on Feynman graphs), $P_G$ is well-defined.  Then, we call $P_G$ the \textbf{period of $\mathbf{G}$}.  Furthermore, primitivity and logarithmic divergence gives necessary and sufficient conditions for the convergence of $P_G$  (see Proposition $5.2$ in ~\cite{motives}), and so we will simply call this condition \emph{primitivity}.  More formally, in terms of graphs we can define primitivity (with logarithmic divergence included) as follows:

\tpoint{Definition} \label{def:primitive}
\statement{
  A graph $G$ is \textbf{primitive} if:
  \begin{itemize}
    \item $n = 2\ell$; where $n = n(G)$ is the number of edges in $G$ and $\ell = \ell(G)$ is the loop number of $G$
    \item every non-empty proper subgraph $\gamma \subset G$ has $n(\gamma) > 2\ell(\gamma)$
  \end{itemize}
}

In ~\cite{numfunct}, one of us outlined a method to calculate some of these periods.  However, in general these periods remain difficult to calculate.  Yet, we would like to understand their properties.  One way forward is through studying the properties of the underlying graphs and manipulating these graphs to find period symmetries.  Another method is through studying related invariants that are easier to work with but still can capture some important information from the period and the graphs, such as the $c_2$~invariant.

There is an interplay between both these methods in the sense that we would also like to find symmetries on graphs that may not preserve the period but some other related invariant.  From the other direction, we would also want any related invariant to preserve some, or ideally all, of the period symmetries found through studying the underlying graphs.

Note that the term "period" comes from algebraic geometry:  Looking at $P_G$ in it's parametric form, if it exists, $\Psi_G$ is simply a polynomial in variables $\alpha_e$ with integer coefficients.  Thus $\Psi_G^{-2}$ is a rational function with $P_G$ a number arising as its integral over $\alpha_e \geq 0$.  That is, $P_G$ is a period as defined by Kontsevich and Zagier ~\cite{Ko-Za} and in the same sense as how multiple zeta values are periods. 

\bpoint{Completion}\label{SS:completion}%
In $\phi^4$-theory, as every logarithmically divergent graph has $4$ external legs, we can uniquely "complete" any such graph $G$ by adding a new vertex connected to all the external edges, giving us a 4-regular graph which is connected if the original graph is connected (and sometimes even if not).
We call this 4-regular graph the $\textbf{completion}$ of $G$.

\begin{figure}[ht]
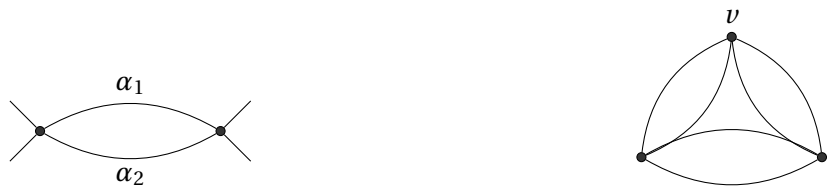

  \centering
  \begin{subfigure}[b]{.5\linewidth}
    \centering
    \GrphOne%
    \caption{A primitive one-loop Feynman graph.}
    \label{fig:G1}
  \end{subfigure}%
  \begin{subfigure}[b]{.5\linewidth}
    \centering
    \CmplOne%
    \caption{The unique completion of (a).}
    \label{fig:Comp1}
  \end{subfigure}%
  \caption{The primitive graph of one loop and its completion.}
  \label{fig:l1}
\end{figure}

Conversely, given a (connected) 4-regular graph $G$, we can delete a vertex $v$ to get a logarithmically divergent $\phi^4$ graph $G - v$.  We call $G - v$ a \textbf{decompletion} of $G$.  Note that you can get non-isomorphic decompletions of the same 4-regular graph.  When the choice of decompletion is not important (either because we are in a case where all the decompletions are isomorphic, or we are interested in a quantity that is invariant under the choice of decompletion), then we will write $\widetilde{G}$ to represent any choice of decompletion of $G$.

We need a notion of primitive for these 4-regular graphs such that by removing any vertex, we stay within primitive 4-point graphs in $\phi^4$-theory (for which the period is well-defined).

\tpoint{Definition} \label{def:completedprim}
\statement{
  A 4-regular graph $G$ with $\geq 3$ vertices is called \textbf{completed primitive} if the only way to split $G$ into multiple connected components with 4 edge cuts is to separate off a vertex, that is there are only trivial 4 edge cuts. In other words, $G$ is internally 6-edge connected.  In this case, we say that $G$ has loop order $\ell$ if $G - v$ has loop order $\ell$ for any vertex $v$.
}

\tpoint{Proposition}\label{prim} (Proposition $2.6$ in ~\cite{census})
\statement{
  Let $G$ be a 4-regular graph and $v$ any vertex in $G$. \\
  Then $G$ is completed primitive if and only if $G - v$ is primitive.
}

\begin{figure}[ht]
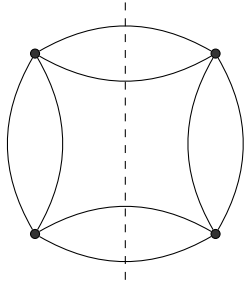
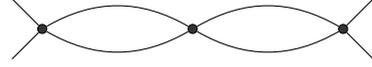

\centering
  \begin{subfigure}[t]{.5\linewidth}
    \centering
    \CmplTwoB%
    \caption{A 4-regular graph on 4 vertices.}
  \end{subfigure}%
  \begin{subfigure}[t]{.5\linewidth}
    \centering
    \GrphTwoB%
    \caption{Decompletion of the 4-regular graph.}
  \end{subfigure}%
  \caption{A non primitive graph and its completion. Notice the trivial 4 edge cut condition in Definition~\ref{def:completedprim} is violated so (a) is not completed primitive. Primitivity (Definition~\ref{def:primitive}) is violated in (b) because (b) contains the primitive one-loop graph (Figure~\ref{fig:G1}) as a subgraph.}
\end{figure}

In ~\cite{census} it is proved that the period is completion invariant.  That is, any two decompletions of the same 4-regular completed primitive graph have the same period:

\tpoint{Theorem} (Theorem $2.7$ of ~\cite{census}) \label{periodcomp}
\statement{
  Let $G$ be a 4-regular completed primitive graph of loop order $\ell$.
  The period of $G - v$ for any vertex $v$, is the same for all choices of $v$.
}     
  
To emphasize the invariance of the previous theorem we will often write $P_{\widetilde{G}}$; we can think of this period as belonging to $G$ itself, and so sometimes for a completed primitive graph $P_G$ is defined to mean $P_{G-v}$, but we will not do this so as to avoid confusion.
  
This completion symmetry tells us that rather than looking at periods of 4-point graphs, we can instead focus on 4-regular graphs. Note that completion considerably reduces the number of relevant graphs at each loop order $\geq 5$.

\tpoint{Example} (Period calculation of Figure~\ref{fig:G1})

Consider the (unique) primitive graph with one loop $G$ (Figure~\ref{fig:G1}). Using Eq.~\eqref{parametric_period}, the parametric space representation of the period, we get
\begin{align*}
  P_G
    &= \bigintsss_{\; 0}^{\infty} \mathrm{d} \alpha_2 \,\frac{1}{\left. \Psi_{G}^2 \right\rvert_{\alpha_1=1}}
    = \bigintsss_{\; 0}^{\infty} \mathrm{d} \alpha_2 \,\frac{1}{\left. (\alpha_1 + \alpha_2)^2 \right\rvert_{\alpha_1=1}} \\
    &= \bigintsss_{\; 0}^{\infty} \mathrm{d} \alpha_2 \,\frac{1}{(1 + \alpha_2)^2}
    = - \left. \frac{1}{1+\alpha_2} \right\rvert_0^{\infty} = 1.
\end{align*}

\bpoint{The $\mathbf{c_2}$ invariant}\label{SS:c2}%
Looking at the period in its parametric form, notice we are integrating over the denominator $\Psi_G^2$, which is just a polynomial in $\abs{E(G)}$ variables.  In particular, in order to understand and characterize properties of the period, we need to understand the structure of $\Psi_G$.

This motivates the study of the zeros of $\Psi_G$ and the polynomials (denominators) that arise after several steps of integration.  In ~\cite{fq}, one of us introduced the following arithmetic invariant and proved it is well-defined:

\tpoint{Definition} (Theorem $2.9$ of ~\cite{fq}) \label{def:c2}
\statement{
  Let $q$ be a prime power and $\mathbb{F}_q$ the finite field with $q$ elements.
  Let $G$ be a connected graph with at least $3$ vertices.
  Then the \textbf{$\mathbf{c_2}$-invariant} of $G$ at $q$ is
  \[ c_2^{(q)}(G) \equiv \frac{\left[\Psi_G\right]_q}{q^2} \mod q \]
  where $\left[\Psi_G\right]_q$ is the number of zeros of $\,\Psi_G$ in $\mathbb{F}_q^{\abs{E(G)}}$. \\
  Denote by $\mathbf{c_2(G)}$ the sequence of $c_2^{(q)}(G)$ for all prime powers $q$.
}

The $c_2$ invariant is relatively easy to calculate, compared to the period, at least for low loop orders or small primes $q=p$, and still encapsulates many of the properties of the period and its underlying graph. For further details we refer the reader to~\cite{brown2014properties}.  However, whether the completion symmetry holds for the $c_2$ invariant is still unknown.

\tpoint[(Brown and Schnetz; $\mathbf{c_2}$ completion)]{Conjecture} (Conjecture $4$ of ~\cite{k3}) \label{c2decompconj} \\
\statement[eq]{
  Let $G$ be a connected 4-regular graph and $v$ and $w$ be vertices of $G$.
  Then
  \[ c_2(G - v) = c_2(G - w). \]
}

Optimistically, there has been some progress in the special case of $q = 2$ (see ~\cite{specialc2}). It is also known when there is a double triangle (defined in Section~\ref{S:doubletriangle}), we can reduce the problem to smaller graphs. 

\bpoint{Graph polynomials}\label{SS:poly}%
Let $G$ be a connected graph.  Recall that for the graph polynomial of $G$, we associate a Schwinger parameter $\alpha_e$ to each edge $e$ and get
\[ \Psi_G = \sum_{T \subseteq G} \prod_{e \not\in E(T)} \alpha_e \]
where the sum runs over all spanning trees of $G$. 

To represent this polynomial as a determinant of a matrix, we first define the following:

\tpoint{Definition}
\statement{
  Given a connected graph $G$, choose an arbitrary orientation on the edges.
  Let $\cE_G$ be the $\abs{V(G)} \times \abs{E(G)}$ signed incidence matrix, with any one row (corresponding to a vertex) removed.
  Let $A$ be the diagonal matrix of $\alpha_e$ for $e$ in $E(G)$, in the same order as the
  columns of $\cE_G$.
  Then we define the \textbf{expanded Laplacian} of $G$ to be
  \[ M_G = \left[
    \renewcommand\arraystretch{1.5}
    \begin{array}{c|c}
      A & {\cE_G}^T \\
      \hline
      -\cE_G & 0 \\
    \end{array}
    \right].
  \]
}

While this matrix is not well-defined as it depends on the choice of row removed in $\cE_G$ as well as the choice of orderings and orientation, we have
\[ \Psi_G = \det\left(M_G\right) \]
for any such choice.  We can then define the following:

\tpoint{Definition}\label{def:dogdsonpoly}
\statement[eq]{
  Let $I$, $J$, and $K$ be subsets of edges of $G$ such that $\abs{I} = \abs{J}$.
  Without restriction we can assume $I\cap K=J\cap K=\emptyset$.
  Denote $M_G{(I, J)}_K$ the matrix obtained from $M_G$ by deleting rows indexed by $I$ and columns indexed by $J$, and setting $\alpha_e = 0$ for $e \in K$.
  Then the \textbf{Dodgson polynomial} is defined to be
  \[ \Psi_{G, K}^{I, J} = \det M_G(I, J)_K. \]
}

This polynomial is well-defined up to sign which depends on which choice of $M_G$ is used
(we keep a choice of $M_G$ fixed from now on).  When the graph $G$ is clear from the context, we will drop the subscript of $G$.  

We will need the following properties of Dodgson polynomials, also found in Sections 2.2, 2.3 of ~\cite{periods}, though care must be taken with the signs.

\tpoint{Proposition}\label{contdel}
\statement{
  Let $e$ be an edge in $G$. Define $\sigma_e = (-1)^{n_e}$ where ${n_e = |\{i \in I : i < e\}| + |\{j \in J : j < e\}|}$ and $i < e$ means $i$ has a smaller index than $e$ in $M_G$.
  \begin{itemize}
    \item Deleting the edge $e$ corresponds to deleting the row and column corresponding to $e$ in $M_G(I, J)_K$:
      \[ \Psi_{G\setminus e, K}^{I,J} = \det M_G(I \cup e, J \cup e)_K = \Psi_{G, K}^{I \cup e, J \cup e} \]
    \item Contracting the edge $e$
    (keeping multiple edges and self loops) corresponds to setting the variable $\alpha_e$ to 0 in $M_G(I, J)_K$:
      \[ \Psi_{G/e, K}^{I,J} = \sigma_e\det M_G(I, J)_{K \cup e} = \sigma_e \Psi_{G, K \cup e}^{I, J} \]
  \end{itemize}
  Thus we have
  \[ \Psi_{G, K}^{I, J} = \sigma_e (\Psi_{G \setminus e, K}^{I, J} \alpha_e + \Psi_{G/e, K}^{I,J}). \]
  That is, Dodgson polynomials satisfy a deletion-contraction relation.
  It also follows that by passing to a minor of $G$, we can assume $I \cap J = K = \emptyset$ as
  \[ \Psi_{G, K}^{I, J} = \pm\Psi_{G', \emptyset}^{I', J'}\]
  where $G' = G \setminus (I \cap J) / (K \setminus (I \cap J))$,
  $I'=I\setminus (I \cap J)$, and $J'=J\setminus (I \cap J)$.
}

\begin{proof}
The deletion relation $\det M_G(I \cup e, J \cup e)_K = \det M_{G\setminus e}(I,J)_K$ holds up to sign because the matrices are the same.

Take the determinant of $M_G(I,J)_K$ by cofactor expansion along the row or column where $\alpha_e$ resides. If $\alpha_e$ is in row and column $k$ in $M_G$, then $\alpha_e$ is in row $k - |\{i \in I : i < e\}|$ and column ${k - |\{j \in J : j < e\}|}$ in $M_G(I,J)_K$. The cofactor corresponding to $\alpha_e$ has a factor of $(-1)^{n_e} = \sigma_e$. We get that
\[\Psi^{I,J}_{G,K} = \sigma_e \alpha_e \det M_G(I\cup e,J\cup e)_K + \det M_G(I,J)_{K \cup e}.\]

We now prove the contraction relation. First, we note that the incidence matrix $\cE_{G/e}$ is obtained from $\cE_G$ by applying row operations until there is only a single non-zero $1$ or $-1$ entry left in the $k$th column, and then removing the row and column corresponding to that non-zero entry.  In $M_G(I,J)_{K\cup e}$, after reducing the row and column of $\alpha_e$ in $M_G(I,J)$ to a single non-zero element, take the cofactor expansion along the row (or column) containing that single non-zero element. The resulting submatrix is $M_{G/e}(I,J)_K$. The cofactor expansion yields a sign of $(-1)^{n_e}$ by a similar argument as above. This proves both the contraction and deletion-contraction relations.
\end{proof}

Using the matrix tree theorem:

\tpoint{Lemma}\label{matrixtree}
\statement{
  Let $U$ be a subset of edges of $G$ such that ${\abs{E(G \setminus U)} = \ell(G) = \abs{E(G)} - \abs{V(G)} + 1}$. \\
  Let $\cE_G(G\setminus U)$ denote the square $(\abs{V(G)} - 1) \times (\abs{V(G)} - 1)$ matrix obtained from $\cE_G$ by deleting the columns indexed by the edges of $G \setminus U$ (recall that $\cE_G$ already has one row removed).
  Then
  \[ \det \cE_G(G\setminus U) =
  \begin{cases}
    \pm 1 & \text{if } U \text{ is a spanning tree of } G \\
    0 & \text{otherwise}
  \end{cases}\]
}
we get the following:

\tpoint{Proposition}\label{dodgsonsum}
\statement{
  Suppose $I \cap J = K = \emptyset$.
  Then we have
  \[ \Psi_{G, \emptyset}^{I, J} = \sum_{U \subset G \setminus (I \cup J)}  \det\left( \cE_G( G \setminus (U \cup I)) \right) \det\left( \cE_G( G \setminus (U \cup J)) \right)\prod_{u \not\in U} \alpha_u\]
  where the sum runs over all subgraphs $U$ such that $U \cup I$ and $U \cup J$ are both spanning trees in $G$.
}

One specific combination of Dodgson polynomials of importance is the 5-invariant. 

\tpoint{Definition}\label{5invdef}
\statement{Given edges $1,\,\dots\,,5$ for a graph $G$, define the \textbf{5-invariant} of G,  $^5\Psi_G(1,\,\dots\, ,5)$ as
    \[
    ^5\Psi_G (1,\,\dots\, ,5) = \pm \det \left(
    \renewcommand\arraystretch{1.25}
    \begin{array}{cc}
        \Psi_5^{12,34} & \Psi_5^{13,24} \\
        \Psi^{125,345} & \Psi^{135,245}\\
    \end{array}
    \right).
   \]
}
The 5-invariant is defined up to overall sign.  Furthermore, permuting the order of the edges $1,\,\dots\, ,5$ only changes the sign of $^5\Psi_G(1,\,\dots\, ,5)$, see Lemma 87 in ~\cite{periods}. 

\bpoint{Denominator reduction}\label{SS:denom}%
Given a graph G and a sequence of edges $e_1,\,\dots\, ,e_{|E(G)|}$ we define
\[D^5_G(e_1,\,\dots\, ,e_5) =  {}^5\Psi_G(e_1,\,\dots\, ,e_5).\]
 To define $D^n_G(e_1,\,\dots\, ,e_n)$ for $n > 5$, we do so recursively. Suppose $D^n_G(e_1,\,\dots\, ,e_n)$ is a polynomial in variables $\alpha_{n+1},\alpha_{n+2},\,\dots\, ,\alpha_{|E(G)|}$. Then if $D^n_G(e_1,\,\dots\, ,e_n)$ factors as
 \[D^n_G(e_1,\,\dots\, ,e_n) = (A\alpha_{n+1}+B)(C\alpha_{n+1}+D),\]
 we define
\[D^{n+1}(e_1,\,\dots\, ,e_{n+1}) = \pm(AD-BC).\]
This process ends when $D^{n+1}_G = 0$ or $D^n_G$ cannot be factored.

Note that after $n = 5$, the ability to factor the polynomial in the desired form may depend on the sequence of edges chosen, so the process may terminate sooner for some edge orderings compared to others. However, $D^n_G(e_1,\,\dots\, ,e_n)$ is independent of the choice of ordering of $e_1, e_2, \ldots, e_n$ for every order for which it is defined.  This process is called \textbf{denominator reduction}. We call $D^n_G(e_1,\,\dots\, ,e_{n})$ the \textbf{$\mathbf{n}$-invariant} and also refer to them with the notation $^n\Psi_G(e_1,\,\dots,\,e_n)$. They are defined up to overall sign.

The name denominator reduction comes from the fact that $D^n_G$ actually arises as the denominator after integrating ("reducing") the indicated $n$ edge variables from the period of $G$ (see Section 10 in~\cite{periods}).  Thus we can also define $D^n_G$ for $n < 5$, however we have to sacrifice the invariant aspect of it.  That is, $D^n_G$ now depends on the edge orderings up to $n$ and has many distinct possible choices.  However each of these choices leads to the $5$-invariant under the denominator reduction process defined above, so from the point of view of any quantity or property which is unchanged under denominator reduction these different $D^n_G$ are equivalent.  Theorem~\ref{denomredc2} is an example of this.

We take $D^n_G(e_1,\,\dots,\,,e_n)$ for $n =3$ and $n=4$ to be as follows.
$D^3_G(i,\,j,\,k)$ for distinct edges $i$, $j$, and $k$ is defined to be
\begin{equation}\label{D3}
    D^3_G(i,\,j,\,k) = \pm\Psi^{ik,jk}\Psi^{i,j}_k
\end{equation}
and $D^4_G(i,\,j,\,k,\,l)$ for distinct edges $i$,  $j$, $k$, and $l$ is defined to be
\[
    D^4_G(i,\,j,\,k,\,l) = \pm\Psi^{ij,kl}\Psi^{ik,jl},
\]
both defined up to sign.  With these definitions, the $D^3_G$ is the denominator after reducing edges $i$, $j$, and $k$ in that order, however it depends on the order of $i$, $j$, $k$ for more than just sign, typically yielding truly different polynomials.  After four integrations, the integrand can be written as a sum over three terms, one with each of the three $D^4_G$s built from the four integrated edges as its denominator.  However, applying the denominator reduction algorithm to any one of them gives the 5-invariant, and so for denominator reduction invariant properties, it is sufficient to consider any one of the $D^4_G$s.

This notion of a higher invariant is useful for the calculation of $c_2$ invariants by the following theorem.

\tpoint{Theorem} (Theorem $29$ of ~\cite{k3} with the statement and proof of Corollary 28 of ~\cite{k3} for $n<5$) \label{denomredc2} 
\statement[eq]{
  Let $G$ be a connected graph with $2\ell \geq \abs{E(G)}$ and $\abs{E(G)} \geq 5$. 
  Suppose that $D^n_G(e_1,\,\dots\, ,\, e_n)$ is the result of the denominator reduction after $3 \leq n < \abs{E(G)}$ steps.
  Then
    \[ c_2^{(q)}(G) \equiv (-1)^n \left[ D^n_G(e_1,\,\dots\, ,\, e_n)\right]_q \mod q.\]
}

For special configurations denominator reduction is particularly efficient.  To utilize this the following identities on Dodgson polynomials will be useful:

\tpoint{Proposition} \label{dodgsonidentities}
\statement[eq]{
  \begin{itemize}
      \item Suppose $\{i,j,k\}$ forms a triangle in $G$.
        \begin{itemize}
          \item If $\{i,j,k\} \subseteq (K \cup I) \setminus J$ then $\Psi_{G,K}^{I,J} = 0$.
          \item If $\{i,j\} \in (K \cup I)\setminus J$ with $k \not \in I \cup J \cup K,$ then $\Psi_{G,K}^{I,J}$ is divisible by $\alpha_k$.
        \end{itemize}
      \item Suppose $\{i,j,k\}$ is a cut set in $G$.
        \begin{itemize}
          \item If  $\{i,j,k\} \subseteq I$ then $ \Psi_{G,K}^{I,J} = 0$.
          \item If $\{i,j\} \subseteq I$, with $k \not \in I \cup J \cup K,$ then $\Psi_{G,K}^{I,J}$ is independent of $\alpha_k$.
          \item If $i\in I\setminus J$ and $\{j,k\} \subseteq J\setminus I$ then $\Psi_{G,K}^{I,J}=\pm\Psi_{G\setminus i/j,K}^{I\setminus i,J\setminus j}=\pm\Psi_{G\setminus i/k,K}^{I\setminus i,J\setminus k}$;
            furthermore if $j,k$ are larger than all other indices of $I$ and $J$ then the signs are all positive.
        \end{itemize}
  \end{itemize}
}

\begin{proof}
The first four points are Proposition 3.19 from ~\cite{yeats2015some}.  For the final point, it suffices to prove the result in the case $K=I\cap J = \emptyset$ by passing to a minor, as described previously.  With this assumption, we wish to compare the edge sets which are spanning trees in both $G\setminus I / J$ and $G\setminus J / I$ with those that are spanning trees in both $(G\setminus i / j) \setminus (I\setminus i) / (J\setminus j) = G\setminus I / J$ and $(G\setminus i/j) \setminus (J\setminus j) / (I \setminus i)$
(see Proposition~\ref{dodgsonsum}). Since $\{i,j,k\}$ is a cut set, $i$ is a bridge in $G\setminus J$ and so the vertex to which $i$ is contracted in $G\setminus J/I$ is a cut vertex.  Similarly $j$ is a bridge in $G\setminus i \setminus (J\setminus j)$ and so the vertex to which $j$ is contracted in $(G\setminus i/j) \setminus (J\setminus j) / (I \setminus i)$ is a cut vertex.  Furthermore, since $\{i,j,k\}$ is a cut set, in both cases, the two subgraphs joined at the cut vertex are the minors of $G$ coming from the two components of $G\setminus \{i,j,k\}$ after deleting $J\setminus \{j,k\}$ and contracting $I\setminus i$ (though the vertices at which the subgraphs are joined differ in general between the two cases).  Thus the spanning trees are the same in both cases and so $\Psi_{G,K}^{I,J}=\pm\Psi_{G\setminus i/j,K}^{I\setminus i,J\setminus j}$.  The same argument with $j$ and $k$ swapped gives the final equality.

To see the relative signs between the terms, fix a set of edges contributing a nonzero term to these Dodgsons.  This edge set determines two full rank submatrices of $\mathcal{E}_G$.  The sign of the term corresponding to this edge set in $\Psi_{G,K}^{I,J}$ is the product of the determinants of these submatrices of $\mathcal{E}_G$.  The analogous term in $\Psi_{G\setminus i/j,K}^{I\setminus i,J\setminus j}$ has sign the product of determinants of the analogous submatrix of $\mathcal{E}_{G\setminus i/j}$.  Since $i\in I$, $i$ has no effect on this sign.  The contraction of $j$ can be implemented at the level of matrices by using column operations until column $j$ has only one nonzero entry and then removing the row and column of that entry.  We can do this with only the column operation of adding a multiple of a column to column $j$ and so not affecting any determinant, and we can do it the same way for both matrices.  Then the only effect of contracting $j$ in the determinants between $G$ and $G\setminus i/j$ is the product of the value of the remaining entry of $j$ and the sign this entry gets in cofactor expansion.  Since $k\in J$, and all other entries of $I$ and $J$ have smaller indices, the columns of $j$ is the same column in both matrices and since we used the same column operations in both matrices, both matrices contribute the same sign, hence the sign difference between $\Psi_{G,K}^{I,J}$ and $\Psi_{G\setminus i/j,K}^{I\setminus i,J\setminus j}$ is 1.  Swapping $j$ and $k$ in this argument gives the signs in the final equality.
\end{proof}

From Section 2.3 items (1) and (2) in ~\cite{k3} we have the following statement: If $i\in I$ and $j\in J$ are a double edge or the edges of a 2-valent vertex then
\begin{equation}\label{2valentvertex}
    \Psi_{G,K}^{I,J}=\pm\Psi_{G\setminus i/j,K}^{I\setminus i,J\setminus j}=\pm\Psi_{G\setminus j/i,K}^{I\setminus i,J\setminus j}.
\end{equation}

As we only care about completed primitive graphs $G$, the only 3-edge cut sets are 3-valent vertices, that is when $\{i, j, k\}$ meet at a common vertex.

From this Proposition, we get the notion of "free" factorizations of denominators, in the sense that if two of the three edges in a triangle or a 3-valent vertex are already reduced in the 5-invariant, then using an appropriate ordering of edges, we can reduce the third edge such that there is no constant term (for triangles) or no quadratic term (for 3-valent vertices) in the 5-invariant which leads to a denominator reduction which always factors.

In certain cases, denominator reduction allows us to dramatically reduce the complexity of the polynomials
we are working with. This is especially useful for computing $c_2$ invariants.

A key result that we will use in the computation of $c_2^{(p)}(G)$ is the following. See Section 2 in ~\cite{ax1964zeroes} or Lemma 2.6 in ~\cite{yeats2016few} for a proof. 

\tpoint{Theorem} (Corollary of Chevalley-Warning theorem) \label{cw}
\statement{
    Let $F$ be a polynomial of degree $N$ in $N$ variables, $x_1,\,\dots\, ,x_N$, with integer coefficients. Then we have
    \[ \text{coefficient of } x_1^{p-1}\,\cdots\, x_N^{p-1} \text{ in } F^{p-1} \equiv (-1)^{N-1}[F]_p \mod p. \]
}
Note that $D^n_G(e_1,\,\dots\, ,e_n)$ satisfies the criterion for Theorem~\ref{cw} for $n \geq 5$. 

\bpoint{Graphs data}\label{SS:data}%
When describing graphs, we use the same convention as in ~\cite{census}. That is, each completed primitive graph will be denoted $P_{\ell,n}$ where $\ell$ is the loop number after decompletion and $n$ is a positive integer which describes the order in which the graphs were generated. Practically $n$ is of not much use other than as a label.

We used the "Periods" file in the arXiv submission of ~\cite{coaction} (an updated version is in ~\cite{Hlogproc}) which contains information about completed primitive $\phi^4$ graphs up to loop order 11, including previously known periods and $c_2$ invariants, as our reference along with additional data from Erik Panzer ~\cite{erik}. We also referred to this list of graphs when implementing the Fourier split and computing $c_2$ invariants at 11 loops. 

\section{The Fourier Split}\label{S:fouriertwist}
First we will start by studying graph transformations corresponding to variable transforms in the integrand, which gives rise to period identities.  These types of symmetries are important as they give equivalence classes of 4-regular graphs where all decompletions of every member of the class have the same period. 

There are currently four known period symmetries: completion, products, planar duality (called the Fourier identity) and the twist.  We will prove a new graph transform that once again preserves the period and arises from the ideas of the Fourier and twist transforms. 

\bpoint{Period symmetries}\label{SS:syms}%
One nice property of 4-regular completed primitive graphs is that they only have vertex-connectivities of 3 or 4.  Note that these graphs have trivial 4 vertex splits (as every vertex has degree 4) but may have that the non-trivial vertex cuts (that is cuts which separate off more than one vertex) are larger.

\tpoint{Definition} \label{def:reducible}
\statement{
    A completed primitive graph is called \textbf{reducible} if it has vertex-connectivity 3.
    Otherwise it is called \textbf{irreducible}.
}

There is a nice product identity for reducible graphs which means we only need to look at irreducible graphs.

\tpoint[(The product identity)]{Theorem} (Theorem $2.10$ in ~\cite{census}) \label{prodsplit}
\statement[eq]{
    A reducible completed primitive graph $G$ is the gluing of two completed primitive graphs $G_1$ and $G_2$ on triangle faces followed by removing the edges of the triangle. 
    The period of $\widetilde{G}$ is thus the product of the periods of $\widetilde{G}_1$ and $\widetilde{G}_2$:
    \[ P_{\widetilde{G}} = P_{\widetilde{G}_1} P_{\widetilde{G}_2}\]
}

From here onward, we consider only irreducible completed primitive 4-regular graphs $G$. 

Used as early as ~\cite{knots} on this particular problem, a natural period identity arises from reinterpreting the Fourier transform taken to get from momentum space to position space as a graph transform. 

Graphically we notice that if $G-v$ is planar, by taking the dual $G'$ of $G - v$, the vertices of $G'$ are marking the cycles of $G - v$.  That is, the momentum space period of $G - v$ is the same as the position space period of $G'$.

\tpoint[(The Fourier identity)]{Theorem} (Theorem $2.13$ and Remark $2.15$ in ~\cite{census})\label{dual}\\
\statement{
  Let $G$ be a 4-regular completed primitive graph.
  Suppose we can make $G$ planar by removing one vertex, say $v$.
  Let $G'$ be the dual of $G - v$.
  If $G'$ can be completed to a 4-regular graph $H$ (i.e. by adding one vertex) then we have
  \[ P_{\widetilde{H}}  = P_{\widetilde{G}} \]
  and $H$ is completed primitive.
  Furthermore $H$ is reducible if and only if $G$ is reducible.
}

In ~\cite{census}, one of us introduced a new transform that is period invariant, called the \textbf{twist}. 

Let $G$ be a 4-regular graph.  Suppose there exists a separation of $G$ say $\{X_1, X_2\}$ of order 4 (that is $X_1$ and $X_2$ partition the edges of $G$ and the subgraphs they induce share exactly 4 vertices).  By abuse of notation we will also refer to the subgraphs induced by $X_1$ and $X_2$ as $X_1$ and $X_2$.

Suppose $X_1 \cap X_2 = \{a, b, c, d\}$ are the 4 vertices that disconnect $G$.  Now, identify vertices $a$ and $b$ from $X_1$ to $b$ and $a$ from $X_2$ (respectively).  Similarly identify vertices $c$ and $d$ from $X_1$ to $d$ and $c$ from $X_2$ (respectively).  

If the resultant graph $G_0$ is 4-regular then let $H = G_0$.  If not, then assuming it is possible, uniquely swap edges $(a,c)$ and $(b,d)$ or $(a,d)$ and $(b,c)$ to get a 4-regular graph $H$.

\tpoint[(The twist identity)]{Theorem} \label{twist} (Theorem $2.11$ and Remark $2.12$ in ~\cite{census})\\
\statement{
  Let $G$ be as above with $H$ its twist.
  Then $H$ is a 4-regular completed primitive graph with
  \[ P_{\widetilde{H}} = P_{\widetilde{G}}. \]
  Furthermore $H$ is reducible if and only if $H$ is reducible.
}

The idea of transforming only one component of $G$, while keeping the other the same, can be extended to include the Fourier transform.  That is, if possible, taking the dual in some appropriate way of one component of $G$ should give rise to another period identity.

In the following, we prove that this "half-dual" transform, call it the Fourier split, in fact does preserve the period.

\bpoint{Graphical functions}\label{SS:graphfuncs}%
To prove this new period identity, we will need some machinery from the theory of graphical functions, as first developed by one of us ~\cite{graphfuncs}.  This is also the same theory that helped prove the zig-zag conjecture ~\cite{zigzag}.

\tpoint{Definition}\label{def:graphfuncs}
\statement{
  Let $G$ be a graph with three distinguished vertices labelled $0$, $1$ and $z$. 
  We call these three vertices, $\textbf{external vertices}$. \\
  \indent The \textbf{graphical function} of $G$, $f_G$, is defined to be the period of $G$ in position space without integrating over $x_z$, the variable associated with $z$. 
  That is $f_G$ is a function of $x_z$ and
  \[ f_G(x_z) = \bigintsss \prod_{v \neq 0, 1, z \,\in V(G)} \left( \frac{\mathrm{d}^4 x_v}{\pi^2} \right) \,\frac{1}{\displaystyle\prod_{e=(i,j)} \left. (x_i - x_j)^2 \right\rvert_{x_0=0, x_1=\mathbb{1}}}\] \label{f:graphfunc}
  where
  \[ P_G = \bigintsss \frac{\mathrm{d}^4 x_z}{\pi^2} f_G(x_z). \]
}

In general, the power of graphical functions comes from the fact that the symmetry of the integral allows one to consider $f_G$ as a function on the complex plane. For more details we refer the reader to ~\cite{graphfuncs}. Here we merely use results from this perspective. So, in this section it is sufficient to leave $f_G$ as a function of the four-dimensional vector $x_z$.

Like the period, graphical functions can also be represented in momentum space ~\cite{graphfuncs} and in parametric and dual-parametric space ~\cite{graphfuncsparam}. 

There is also a more general version of the graphical function of $G$ which allows for
other distinguished subset of vertices called external vertices, for edge weights $\nu_e$ and for dimensions $D > 2$~\cite{graphfuncs, graphfuncsparam, numfunct}. Definition~\ref{f:graphfunc} uses exactly three external vertices labelled $0$, $1$, and $z$, edge weights 1 and dimension $D = 4$. 

Convergence of graphical functions is handled in Lemma 3.4 of ~\cite{graphfuncs}.

For graphs $G$ with a distinguished set of vertices, we have a slightly modified definition of the \textbf{superficial degree of divergence} of $G$:
\begin{equation}\label{MG}
    m_G = E_G - 2V^{\text{int}}_G
\end{equation}
where $V^{\text{int}}_G$ denotes the number of internal vertices of $G$, that is the number of vertices which are not external.

Like the Fourier identity for the period, we have a similar theorem for graphical functions.  First, we need a slightly modified notion of planarity and dual for Feynman graphs.

\tpoint{Definition} \label{extplanar} (equivalent to Definition 4.1 from ~\cite{graphfuncsparam})
\statement{
  Let $G$ be a graph with three external vertices labelled $0$, $1$ and $z$. 
  Let $G_e$ be the graph obtained from $G$ by adding edges $(0, 1)$, $(0, z)$ and $(1, z)$. 
  Then we say that $G$ is \textbf{externally planar} if and only if $G_e$ is planar. \\
  \indent A \textbf{dual} of $G$ is given by taking a dual of $G_e$ with the faces labelled as such:
  \begin{itemize}
    \item Label the inner face created by edge $(1,\, z)$ by $0$
    \item Label the inner face created by edge $(0,\, z)$ by $1$
    \item Label the inner face created by edge $(0,\, 1)$ by $z$
  \end{itemize}
  and removing the star associated with the dual edges of $(0,1)$, $(0,z)$ and $(1,z)$.
}

The condition of being externally planar is equivalent to $G$ having a planar embedding with $0$, $1$ and $z$ on the same face, which without loss of generality can be the external face.  In the graph theory literature this is sometimes known as \emph{planarity with outer terminals}.  Note that in the cases of primary interest to us, $G_e$ will be 3-connected (though $G$ may not be) and so this notion of dual will be unique, see Remark~\ref{rem fourier twist}.

Using this definition of dual, we have the following theorem:

\tpoint{Theorem}(Theorem 1.9 in ~\cite{graphfuncsparam}) \label{dualgf}
\statement[eq]{
  Let $G$ be a graph with three external vertices $0$, $1$, and $z$ such that $f_G(x_z)$ converges and $m_G = 2$. 
  Let $G'$ be the dual of $G$ as defined in Definition~\ref{extplanar}. 
  Then the graphical functions of $G$ and $G'$ are equal:
  \[ f_G(x_z) = f_{G'}(x_z). \]
}

One interesting note about this theorem is that the proof in fact uses the duality between parametric and dual-parametric space, as opposed to the position and momentum space duality originally used in the Fourier identity for the period. 

\bpoint{The Fourier split}\label{SS:FT}%
Let $G$ be a 4-regular graph.  Let $\{X_1, X_2\}$ be a separation of $G$ such that the intersection of the subgraphs induced by $X_1$ and $X_2$ is a 4-vertex cut of $G$, label them $\{0, 1, z, \infty \}$ (recall that we can arbitrarily set any vertex of $G$ to be $0$, $\mathbb{1}$ or $\infty$, and we use the
labels $1$ for $\mathbb{1}$ and $\infty$ for the decompletion vertex $v$).  The edges between the cut vertices may be in either $X_1$ or $X_2$.

Let $\gamma_1$ and $\gamma_2$ be the subgraphs of $G - \infty$ induced by $X_1$ and $X_2$. Further assume that $\gamma_1$ and $\gamma_2$ are connected and both have vertices which were neighbours of $\infty$.  If $G$ has vertex connectivity 4 then this is automatic, see Remark~\ref{rem fourier twist}.  From the point of view of either $\gamma_1$ or $\gamma_2$ the vertices in the cut are external vertices in the sense that they link outside the subgraph (as well as within it).  In view of this we will call $\{0, 1, z\}$ the external vertices of $\gamma_1$ and of $\gamma_2$.

Suppose $\gamma_1$ has $m_{\gamma_1}=2$ (Eq.~\eqref{MG}) and is externally planar with dual $\gamma'_1$ (as defined by Definition~\ref{extplanar}).  Reattach $\gamma'_1$ to $\gamma_2$ by identifying the corresponding vertices $0$, $1$ and $z$. This transform is illustrated in Figure~\ref{fig:FT}. 

\begin{figure}[ht]
  \centering
  \FourierTwist%
  \caption{Top row: $G$ (left) transforming to $\bar{G}$ (right) by a Fourier split.\\
  Bottom row: Fourier split operation on decompleted $G$ to decompleted $\bar{G}$.\\
  The two components are $\gamma_1$ (right) and $\gamma_2$ (left). \\
  The dashed lines show the edges added before dual and its associated dual edges. \\
  The dotted lines show the identification of vertices.\\
  The white dots show the dual vertices; $v$ is the star to be removed.}
  \label{fig:FT}
\end{figure}

If the resulting graph can be completed to a 4-regular graph $H$ then we have:

\tpoint[(The Fourier split identity)]{Theorem}
\statement{
  Let $G$ be a 4-regular completed primitive graph with $H$ its Fourier split as above.
  Then the graph $H$ is completed primitive with
  \[ P_{\widetilde{H}} = P_{\widetilde{G}}. \]
  Furthermore, $H$ is reducible if and only if $G$ is reducible.
}

\begin{proof}
  Consider the period $P_{\widetilde{G}}$ in position space where $G$ is decompleted at $\infty$.
  We have
  \[ P_{\widetilde{G}}= \bigintsss \frac{\mathrm{d}^4 x_z}{\pi^2} \, f_{\gamma_1}(x_z) \; f_{\gamma_2}(x_z).\]
  Because G is completed primitive $P_{\widetilde{G}}$ exists and $f_{G-\infty} = f_{\gamma_1}f_{\gamma_2}$ exists.  This implies the convergence conditions on all subgraphs of $G-\infty$ and so in particular $f_{\gamma_1}$ and $f_{\gamma_2}$ each exist. 
  Moreover, $\gamma_1$ is externally planar with $m_{\gamma_1}=2$, hence $f_{\gamma_1}(x_z)=f_{\gamma'_1}(x_z)$ by Theorem~\ref{dualgf}. Substitution in the above equation gives $P_{\widetilde{H}}$. Because $P_{\widetilde{H}}$ is finite, by the discussion before Definition~\ref{def:primitive}, $H$ is completed primitive.
  
  It remains to prove the final statement.  Because a Fourier split of $P_{\widetilde{H}}$ along $\{0,1,z,\infty\}$ leads back to $G$ it is sufficient for the final statement to prove that if $G$ is reducible then $H$ is reducible.  Suppose, then, that $G$ is reducible, that is, $G$ has a 3-vertex cut.  
  
  If the 3-vertex cut of $G$ is completely on the $X_2$ side of the original 4-separation then the cut trivially survives the Fourier split and so $H$ is also reducible.  If the 3-vertex cut of $G$ is completely on the $X_1$ side, then either $\gamma_1$ has a 2-vertex cut with $0,1,z$ on the same side of the cut or $\gamma_1$ has a 3-vertex cut with $0,1,z$ and the vertices which connected to $\infty$ all on the same side of the cut. 
  
  Note for any planar graph if $\{X_1, X_2\}$ is a separation of order $k$ then the same sets of edges, viewed now as sets of edges in the dual is a separation of the same order and the separation vertices in the dual correspond to the facial cycles involving both parts of the separation (and the separation vertices in the original).  This is an elementary graph theory observation and also can be seen as a consequence of the fact that the connectivity function of a matroid is invariant under duality.
  
  \begin{figure}
      \centering
      \includegraphics{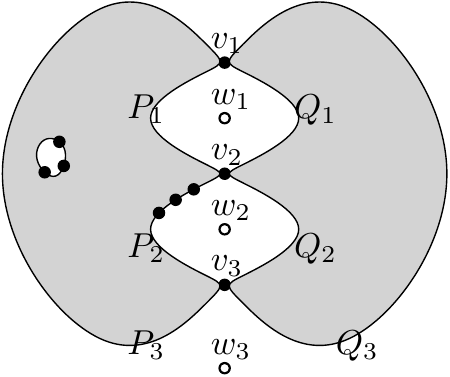}
      \caption{Illustration of how $0$, $1$, and $z$ behave relative to a small cut, along with the notation used in the proof.  If $0$, $1$ and $z$ are on the same side of the cut then they either lie in a face which is not a facial cycle around the cut vertices in the dual (represented by the three black dots on a face on the left) or they lie on the same side of the same facial cycle around a cut vertex of the dual (represented by the three black dots on $P_2$).  In either case $0$, $1$, and $z$ in the dual are also on the same side.}
      \label{fig:cut to dual}
  \end{figure}
  
  Suppose $\gamma_1$ has a $k$-vertex cut (for us $k=2,3$ though the observations below hold for all $k$) with $0,1,z$ on the same side of the cut. Let $v_1, v_2, \ldots, v_k$ be the vertices of the cut and $w_1, w_2, \ldots, w_k$ the vertices of the corresponding cut in $\gamma_1'$ as illustrated in Figure~\ref{fig:cut to dual}.  Consider the facial cycles around the $w_i$, these give pairs of paths $P_i, Q_i$ such that $P_i, Q_i$ gives the facial cycle around $w_1$ and $P_i$ and $Q_i$ are both paths from $v_i$ to $v_{i+1}$ (where $v_{k+1}=v_1$), and where the $P_i$ are on the side containing $0,1,z$ and the $Q_i$ on the other side.  Since $\gamma_1$ is externally planar, $0,1,z$ are on the same facial cycle, so either they are on none of the $P_i$ or $Q_i$, or they are all three on the same $P_i$.  In either case, the corresponding vertices $0,1,z$ in $\gamma_1'$ lie on same side of the $k$ vertex cut, namely the side corresponding to the side they lie on in $\gamma_1$.
   
  If $\gamma_1$ has a 2-vertex cut with cut vertices $\{v_1, v_2\}$ and with $0,1,z$ on the same side of the cut, then the observation of the previous paragraph is sufficient to give that $H$ is reducible. 
  
  Now suppose $\gamma_1$ has a 3-vertex cut with $0,1,z$ and the vertices which connected to $\infty$ on the same side of the cut.  By the argument of the previous paragraph $0,1,z$ are also on the same side of the corresponding 3-vertex cut in $\gamma_1'$.  It remains only to check that the vertices which will be completed in going from $\gamma_1'$ to $H$ are all on the same side of the 3-vertex cut at $0,1,z$. $m_{\gamma_1}=2$ along with the 4-regularity of $G$ guarantees that by adding some choice of two edges between $0$, $1$ and $z$ we can convert $\gamma_1$ into a 4-point $\phi^4$ graph, call it $\eta$.  The graph $\eta$ is also planar has dual $\eta'$ which is $\gamma_1'$ but with the same two additional edges between $0,1,z$.  Since $\hat{G}$ is also 4-regular, $\gamma_1'$ is also a 4-point $\phi^4$ graph.  The 3-vertex cut remains a 3-vertex cut in $\eta$ and all the 3-valent vertices of $\eta$ are on the same side of the cut, so by the irreducibility part of Theorem~\ref{dual} the 3-valent vertices of $\eta'$ are also on the same side of the 3-vertex cut.  Also, in both $\eta$ and $\eta'$ at least one of the 3-valent vertices is one of $0,1,z$. Therefore, removing the two extra edges of $\eta'$, we see that all 3-valent vertices of $\gamma_1'$ are on the same side of the cut as $0$, $1$, $z$. Thus $H$ is reducible.
  
  \begin{figure}
    \centering
    \includegraphics{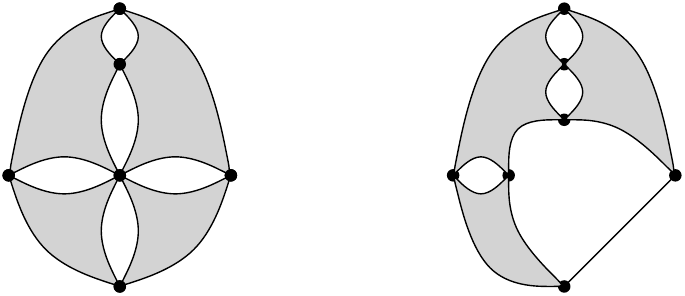}
    \caption{A diagram of how the seventh array configuration looks as a graph along with an example for the fourth array configuration showing how a trivial corner can appear.  Note that trivial corners may also occur for corners with more than two neighbouring vertices.}
    \label{fig:array egs with blobs}
  \end{figure}
  
  Finally, if the 3-vertex cut of $G$ is mixed between $X_1$ and $X_2$ then we claim that we can always find another 3-vertex cut which is entirely in $X_1$ or $X_2$.  To see this proof, we need to consider how the vertex cuts interact.  We have the original 4-vertex cut of $G$ with cut vertices $\{0,1,z,\infty\}$ as well as the supposed 3-vertex cut; let $Y_1, Y_2$ be the separation associated to the 3-vertex cut.
  This partitions the vertex set of $G$ into 9 pieces (some of which may be empty), those vertices in the subgraph induced by $X_1 \cap Y_1$ but not in either cut; those in the 4-vertex cut and in $Y_1$, but not also in the 3-vertex cut; those in $X_2 \cap Y_1$ but not in either cut, and so on.  To visualize these interactions\footnote{We learned this approach from Matt DeVos.}, we will build small arrays as follows:
  \[
    \begin{array}{c|c|c}
         & a & \\
         \hline
         b & c & d \\
         \hline
         & e & 
    \end{array}
  \]
  where $a+c+e = 4$ and $b+c+d=3$. The entries in the arrays give the number of vertices in each of the 9 pieces. The 4-vertex cut runs down the middle, the 3-vertex cut runs across the middle.  The corners of the array are the four sets of vertices not involved in the cuts, and we do not need to record how large these sets are, though we should keep in mind that they may be empty.  In the case that a corner vertex set is empty, there may still be a corresponding subgraph, but it will consist only of edges connecting vertices counted in the neighbouring entries of the array; we'll call this a trivial corner or trivial cut.  See Figure~\ref{fig:array egs with blobs} for examples.  If we have a corner of the array where the three orthogonally and diagonally adjacent entries sum to 3 or less then we have a 3-vertex cut which is entirely on one side of both of the other cuts; we can guarantee it is nontrivial if we can guarantee at least one vertex in the corner set.  Now we simply enumerate possibilities up to symmetry.  Note that not all possibilities can occur in primitive $\phi^4$ graphs.
  \[
      \begin{array}{c|c|c}
         & 3 & \\
         \hline
         2 & 1 & 0 \\
         \hline
         * & 0 & 
    \end{array}
    \quad 
        \begin{array}{c|c|c}
         & 2 & *\\
         \hline
         2 & 1 & 0 \\
         \hline
         & 1 & 
    \end{array}
    \quad 
        \begin{array}{c|c|c}
         & 4 & \\
         \hline
         2 & 0 & 1 \\
         \hline
         * & 0 & 
    \end{array}
    \quad 
        \begin{array}{c|c|c}
         & 3 & \\
         \hline
         2 & 0 & 1 \\
         \hline
         * & 1 & 
    \end{array}
    \quad 
        \begin{array}{c|c|c}
         & 2 & \\
         \hline
         2 & 0 & 1 \\
         \hline
         & 2 & *
    \end{array}
  \]
  \[
      \begin{array}{c|c|c}
         & 3 & \\
         \hline
         1 & 1 & 1 \\
         \hline
         * & 0 & 
    \end{array}
    \quad
        \begin{array}{c|c|c}
         & 2 & \\
         \hline
         1 & 1 & 1 \\
         \hline
         * & 1 & 
    \end{array}
    \quad
        \begin{array}{c|c|c}
         & 4 & \\
         \hline
         3 & 0 & 0 \\
         \hline
         * & 0 & 
    \end{array}
    \quad
        \begin{array}{c|c|c}
         & 3 & * \\
         \hline
         3 & 0 & 0 \\
         \hline
         & 1 & 
    \end{array}
    \quad
        \begin{array}{c|c|c}
         & 2 & \\
         \hline
         3 & 0 & 0 \\
         \hline
         & 2 & *
    \end{array}
  \]
  The corners marked with $*$ all give small vertex cuts and if any of them is trivial then so is one of the original cuts.  Since the cases with 3-vertex strictly on either side of the 4-vertex cut are already dealt with, this completes the proof of the theorem.
\end{proof}   

The condition $m_{\gamma_1}=2$ is often guaranteed in non-trivial cuts.

\tpoint{Lemma}\label{nfaces}
\statement{
  With set up as above, it is always possible to pick a decompletion vertex $\infty$ from a 4-vertex cut such that both subgraphs $\gamma_1$ and $\gamma_2$ have $m_{\gamma_k} = 2$, provided neither subgraph is a star.
}

\begin{figure}[ht]
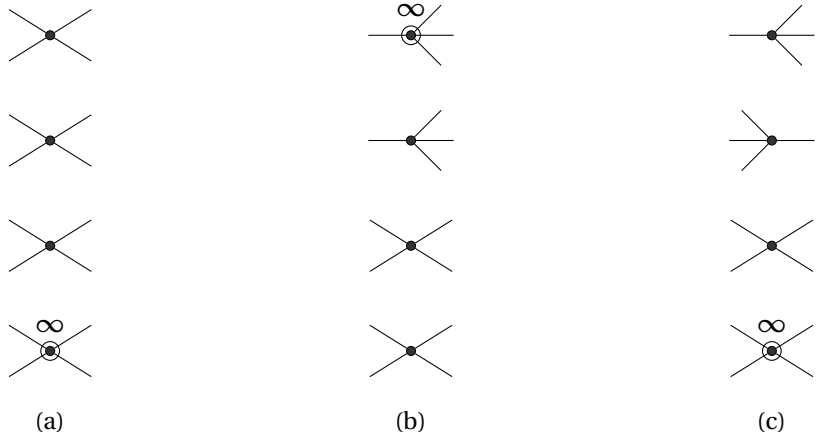

  \centering
  \begin{subfigure}[t]{.3\linewidth}
    \centering
    \ConfigOne%
    \caption{}
    \label{fig:C1}
  \end{subfigure}%
  \begin{subfigure}[t]{.3\linewidth}
    \centering
    \ConfigTwo%
    \caption{}
    \label{fig:C2}
  \end{subfigure}%
  \begin{subfigure}[t]{.3\linewidth}
    \centering
    \ConfigThree%
    \caption{}
    \label{fig:C3}
  \end{subfigure}%
  \caption{Configurations of a 4-vertex cut}
  \label{fig:configs}
\end{figure}

\begin{proof}
  As $G$ is an internally 6-connected 4-regular graph, on both sides of the cut need to be an even number $\geq6$ edges. There are only 3 possible configurations as depicted in Figure~\ref{fig:configs}. We can assume that there are no edges between the external vertices since we can arbitrarily pick which component each of those edges lie in.
  
  Counting half-edges we get
  \[2\abs{E(\gamma_k)} = 4V_{\gamma_k}^{\text{int}}+n_{\neq\infty}-n_\infty,\]
  where $k=1,2$ and $n_\infty$ ($n_{\neq\infty}$) is the number of external edges which
  are (not) connected to $\infty$. The choices of $\infty$ in Figure~\ref{fig:configs} give $n_{\neq\infty}-n_\infty=4$ implying $m_{\gamma_k}=2$.
\end{proof}

\tpoint{Remark}\label{rem fourier twist}
\statement{
  \begin{enumerate}
      \item If $G$ is irreducible then $G-\infty$ is 3-connected.
      In this case we always have vertex-connectivity 3 after adding in the edges to $\gamma_1$; that is the dual is unique: 
      If $\gamma_1$ had a 1-vertex split then there exists an external vertex $0,1,z$ so that $G-\infty$ would have a 2-vertex split. If $\gamma_1$ had a 2-vertex split with an entirely internal component then this split would also split $G-\infty$. So, any 2-vertex split on $\gamma_1$ must have an external vertex and adding $(0,1)$, $(0,z)$ and $(1,z)$ to $\gamma_1$ ensures vertex-connectivity 3.
      \item The Fourier split generalizes the Fourier identity:
      If $\gamma_2$ consists of two edges connected by a vertex then $E_{\gamma_1}-2V_{\gamma_1}^{\text{int}}=E_G-6-2(V_G-4)=2$. Planarity of $\gamma_1$ in the sense of Definition~\ref{extplanar} is equivalent to planarity of $G-\infty$.
      The Fourier split with this setup reproduces the Fourier identity.
      \item As in the cases of the twist and the Fourier identities, this transform is symmetric. In Figure~\ref{fig:FT}, starting from the right (one always has $m_{\gamma'_1}=2$), we use the equivalent definition of dual from Definition 4.1 from ~\cite{graphfuncsparam}. One gets the same sequence of graphs to reach the top left 4-regular graph. Notice that in this example, the Fourier split gives the same result as the twist and the (full) Fourier transform.
      \item The Fourier split---like the twist and Fourier identity---acts on the wider class of non-$\phi^4$ graphs. These graphs may have valence greater than 4 and edges of negative weights
      (numerators in position space). Adding the weights at each vertex still gives 4 but the weights may be $n>4$ times 1 plus $n-4$ times $-1$.
      Convergence of the period is always guaranteed. Within $\phi^4$-theory these graphs are not relevant, but it is conceivable that a sequence of twist and Fourier split relations leads first out of $\phi^4$ and later back into $\phi^4$ again, providing a new relation between $\phi^4$ periods. In this article we did not pursue such a scenario.
  \end{enumerate}
}

Note that when both $\gamma_1$ and $\gamma_2$ are externally planar, applying the Fourier split transform to one side and then the other gives the same result as the (full) Fourier transform as you are now taking a full dual. 

\bpoint{Results}\label{SS:FTresults}%
Implementing the Fourier split in Sage~\cite{sagemath} on completed primitive irreducible graphs up to 11 loops, we get the following results in Table~\ref{tab:comparehepp}, comparing against the Hepp bound (an invariant introduced by E. Panzer, conjectured to be faithful to the period, see ~\cite{hepptalk} for definitions and details) and all currently known period equivalences ~\cite{erik}. The data for the Hepp bound is complete up to 11 loops ~\cite{erik}. 

\begin{table}[ht]
  \centering
  \begin{tabular}{c|c|c}
    $\ell$ & New Identities & Unexplained Identities \\
    \hline
    8 & 0 & 2 \\
    9 & 1 & 6 \\
    10 & 13 & 59 \\
    11 & $53^{*}$ & 229
  \end{tabular}
  \caption{$^{*}$ There are four incomplete unexplained identities.}
  \label{tab:comparehepp}
\end{table}

We have only shown the new results from the Fourier split transform.  All of them preserve the Hepp bound. At loop orders $\ell = 3,\,\dots\, , 7$, all the period equivalences are known and can be explained by Fourier or twist identities. 

In the table, the number of identities is the number of classes of graphs whose periods are (by the new results) or should be (by the Hepp bound) equal.  In particular when such a class has size greater than 2, it is still counted as one identity; for example, if such a class contained three graphs $\{G_1, G_2, G_3\}$ then this is counted as one identity, even though it implies all three of $P_{\widetilde{G}_1}=P_{\widetilde{G}_2}$, $P_{\widetilde{G}_1}=P_{\widetilde{G}_3}$ and $P_{\widetilde{G}_2}=P_{\widetilde{G}_3}$.

A new identity refers to two or more distinct subsets of a class which have proven period equivalences as a result of the Fourier split, and which was not a consequence of (any sequence of) Fourier or twist identities on $\phi^4$ graphs. For most of these new identities, except those that are starred, the entire identity is now proven.  Using these established identities so as to move outside of $\phi^4$ and then back could potentially also give more identities which were not considered here.

An unexplained identity refers to distinct subsets of a class that have the same Hepp bound but for which there are no Fourier, twist, or Fourier split transforms between them that stay within $\phi^4$.  What is meant by an \emph{incomplete unexplained identity} is that the faithfulness conjecture for the Hepp implies some class of three or more graphs should have the same period, while Fourier split calculations along with what was previously known only gave the identity of the periods for some proper subset containing at least two of the graphs.  

Notice that as there are still many unexplained identities, including 2 at $\ell = 8$; there may still possibly be unknown period preserving graph transforms.  Additionally, note that as the Fourier split can only give new results on graphs with non-trivial 4-vertex cuts, the two unexplained identities at $\ell = 8$ were never in contention as the graphs in question have vertex-connectivities of 4 only trivially.  This suggests that if another transform does exist (and the Hepp bound is faithful), it would have to act on higher order vertex cuts or be something entirely new that does not depend on vertex cuts.

An interesting observation is that at all loop orders up to $\ell = 11$, there are also some previously known identities that the Fourier split transform could not capture. Thus this new transform is not sufficient by itself to capture all currently known period identities, even with its connection to both the Fourier and the twist transforms. The Fourier split could also create a direct transform between two graphs that were related through a Fourier or twist identity to a third intermediary graph. 

For a full list of new identities see Appendix~\ref{app:FTtable}, where the incomplete unexplained identities are starred.  All new identities preserve the $c_2$ invariant.  We also include all Fourier, twist and Fourier split results for graphs up to 11 loops in the ancillary files on the arXiv version of this paper.

\section{Double Triangle Reduction and Decompletion}\label{S:doubletriangle}
We now switch our focus to studying a graph transformation called the double triangle (DT) reduction, which does not preserve the period, but does preserve the $c_2$ invariant. While this fact has already been known for primitive 4-point graphs, we will prove that $c_2$ is also preserved by DT reduction in the completed case.

Having this transformation allows us, when studying the $c_2$ invariant, to reduce our problems to smaller graphs and only need to look at those without double triangles.  Additionally, having the completed case is another small step towards proving the $c_2$ completion conjecture (Conjecture~\ref{c2decompconj}).  In particular it settles the $T$ case of ~\cite{specialc2}. 

\bpoint{Double triangle reduction}\label{SS:DTR}%
Suppose a graph $G$ has an edge that is shared by exactly two triangles.  Call this edge $(A,B)$ with triangles $(A,B,C)$ and $(A,B,D)$.

\begin{figure}[ht]
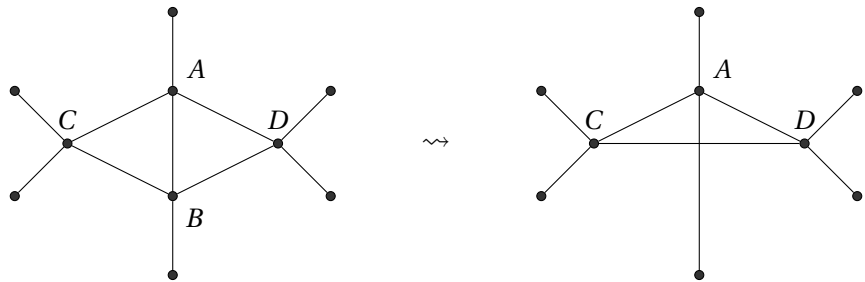

  \centering
  \DTR%
  \caption{Double triangle reduction from left to right \\
  Note: $A, B, C$ and $D$ do not need to be 4-valent}
  \label{fig:DTR}
\end{figure}

Note that the definition does not permit us to do a double triangle reduction if there are three triangles sharing an edge.  This we could call a triple triangle and we only define double triangle reduction for double triangles which are not contained in a triple triangle.  The reason for this restriction is that reducing a double triangle within a triple triangle will cause double edges and is otherwise not well-behaved. (A triple triangle can only be in the complete graph with 5 vertices K5 or in reducible graphs with K5 factors.)

The double triangle reduced graph of $G$ is $G$ with one of the vertices of $(A,B)$, say $B$ replaced with the edge $(C,D)$. If $B$ has a neighbour not in the triangles, then it is now adjacent to the remaining vertex $A$. This is illustrated in Figure~\ref{fig:DTR}. We note that in fact $A, B, C$ and $D$ do not need to be 4-valent (see further discussion below).

\tpoint[($\mathbf{c_2}$ with double triangle)]{Theorem} (Corollary $34$ in ~\cite{k3} using Theorem $35$ in ~\cite{forest}) \label{DTR} \\
\statement[eq]{
  Let $G$ be a primitive-divergent graph in $\phi^4$ and $G'$ be the double triangle reduction of $G$.
  Then
  \[ c_2(G) = c_2(G'). \]
}

Notice that while this result is for primitive graphs, we will prove that double triangle reductions acting on completed graphs also preserve $c_2$. Note that the double triangle reduction is well-defined on completed primitive graphs:

\tpoint{Proposition} (Proposition $2.19$ in ~\cite{census})
\statement{
  A double triangle reduction of a completed primitive graph is completed primitive.
}

\bpoint{Decompletion at a DT vertex}\label{SS:DTRdecomp}%
By Theorem~\ref{DTR}, if a 4-regular completed primitive graph is decompleted at a vertex not adjacent to any of the double triangle vertices, then $c_2$ is preserved.  As a remark, we note that this includes any neighbours of $A, B, C$ or $D$.  The proof used in ~\cite{forest} did not need $C$ or $D$ to be 4-valent, nor was that needed for Theorem $134$ in ~\cite{periods}, which also covers the case where $A$ or $B$ is 3-valent (and in fact is restricted to this case as it deals only with subdivisions of triangles, not general double triangles).  Thus, what is left to show is when the decompletion vertex is a vertex of the double triangle. 

As the double triangle is symmetric, this reduces to two cases: when the decompleted vertex is incident to the shared edge of the triangles (Case 1) and when the decompleted vertex is one of the tips of the triangles (Case 2).  To prove this, we use similar techniques as used in ~\cite{forest} to prove Theorem $35$. 

First, a comment on graphs and graph polynomials.  One way of viewing Dodgson polynomials is through the possible shapes of the underlying graph after any deletions and contractions.  The polynomial can then be thought of as an "intersection" of these graphs where $\cap$ is taken to mean the resulting polynomial of common terms (which are spanning trees in each minor, see also Proposition~\ref{dodgsonsum}).  In the following, we will use this interpretation and notation to show the equality of equations in Dodgson polynomials where the blob is the rest of the graph, usually denoted as $K$.

\begin{figure}[ht]
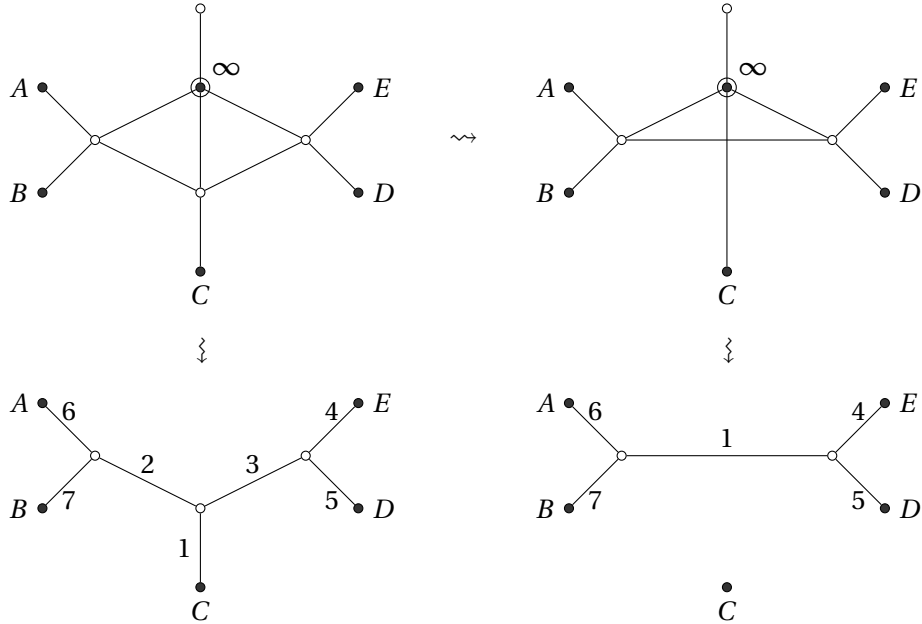

  \centering
  \CaseOne%
  \caption{Decompletion at double triangle vertex Case 1.\\ Double triangle reduction transforms left to right.\\Decompletion at $\infty$ transforms top to bottom.}
  \label{fig:DTR1}
\end{figure}

\tpoint[(Case 1)]{Lemma} \label{DTRCaseOne}
\statement[eq]{
  Let $G$ be a connected 4-regular graph and $G'$ the double triangle reduction of $G$.
  Suppose we decomplete both graphs at the vertex incident to the shared edge remaining after the reduction (see Figure~\ref{fig:DTR1}), denoted $\tilde{G}$ and $\tilde{G'}$ respectively.
  Then
  \[ ^7\Psi_{\widetilde{G}}(2,\, 3,\, 6,\, 4,\, 1,\, 5,\, 7) = \pm\, ^5\Psi_{\widetilde{G'}}(4,\, 5,\, 6,\, 7,\, 1).\]
}

\begin{proof}
  From a 5-invariant of $\widetilde{G}$, as $\{1,\,2,\,3\}$ forms a 3-valent vertex we have (see Definition~\ref{5invdef})
  \[
    \pm^5\Psi_{\widetilde{G}}(2,\, 3,\, 6,\, 4,\, 1)
      = \Psi_1^{23,46}\Psi^{126,134} - \Psi^{123,146}\Psi_1^{26,34}
      = \Psi_1^{23,46}\Psi^{126,134}.
  \]
  Then since $\{2,\,6,\,7\}$ and $\{3,\,4,\,5\}$ also form 3-valent vertices, we get the denominator $D^7_{\widetilde{G}}$ for free:
  \[
    \pm^7\Psi_{\widetilde{G}}(2,\, 3,\, 6,\, 4,\, 1,\, 5,\, 7)
      = \Psi_1^{2357,4657}\Psi_{57}^{126,134}.
  \]
  
  Similarly, from a 5-invariant of $\tilde{G'}$, as $\{1,\,4,\,5\}$ forms a 3-valent vertex, we have
  \[
    \pm^5\Psi_{\widetilde{G'}}(4,\, 1,\, 6,\, 7,\, 5)
      = \Psi^{456,157}\Psi_5^{14,67}.
  \]
  
  From here, using the graphical interpretation of Dodgsons, we get equality automatically by looking at the underlying graphs to the polynomials $\Psi_{G, K}^{I, J}$ viewed as spanning trees at the intersection of $G\setminus I/\{J \cup K\}$ and $G\setminus J/\{I \cup K\}$. In this case both $^7\Psi_{\widetilde{G}}$ and $^5\Psi_{\widetilde{G'}}$ reduce to the following (up to sign):
  \begin{center}\CaseOneDTBlob\end{center}
  While not needed for this proof, we can also rephrase this equality as an equality on spanning forest polynomials of the blobs (see ~\cite{forest}).
  
  To view this equality directly using Dodgson properties, we use contraction-deletion (Proposition~\ref{contdel}) and Eq.~\eqref{2valentvertex} to get
  \begin{align*}
    \Psi_{\widetilde{G},1}^{2357,4657}\Psi_{\widetilde{G},57}^{126,134}&=\Psi_{\widetilde{G}\setminus57/1}^{23,46}\Psi_{\widetilde{G}\setminus1/57}^{26,34}=\Psi_{\widetilde{G}\setminus4567/123}\Psi_{\widetilde{G}\setminus12/357}^{6,4}=\Psi_{\widetilde{G'}\setminus146/57}\Psi_{\widetilde{G'}\setminus1/57}^{6,4} \\
    \Psi_{\widetilde{G'}}^{456,157}\Psi_{\widetilde{G'},5}^{14,67}&=\Psi_{\widetilde{G'}\setminus5}^{46,17}\Psi_{\widetilde{G'}/5}^{14,67}.
  \end{align*}
  Twice applying Eq.~\eqref{2valentvertex} to the first factor for $\widetilde{G'}$ gives $\Psi_{\widetilde{G'}\setminus146/57}$. 
  For the second factor, we use the fact that \{1,\,6,\,7\} is a 3-valent vertex
  and obtain by Proposition~\ref{dodgsonidentities}, $\Psi_{\widetilde{G'}\setminus1/57}^{4,6}$. Comparison with the result for $\widetilde{G}$ proves the lemma.
\end{proof}

\begin{figure}[ht]
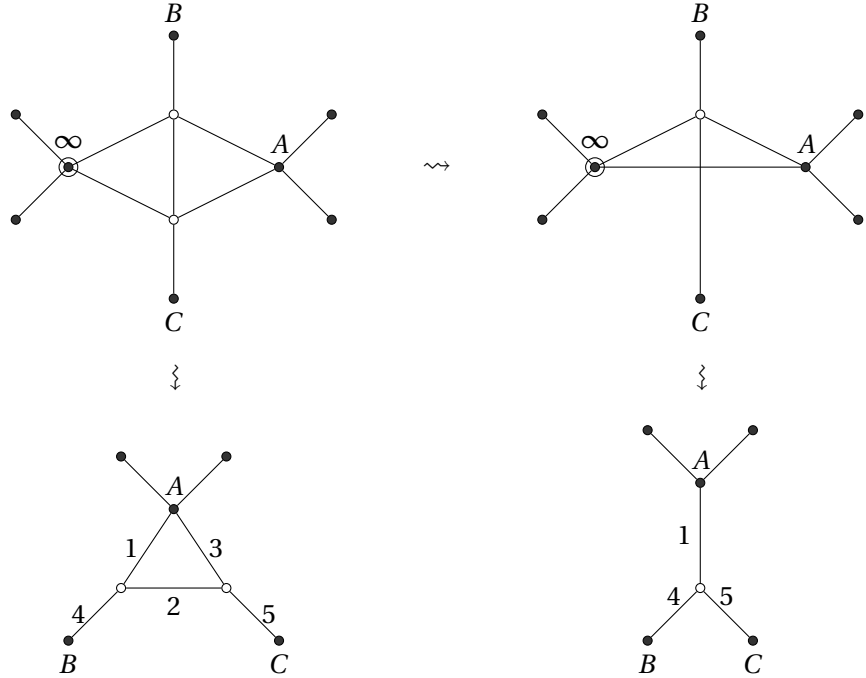

  \centering
  \CaseTwo%
  \caption{Decompletion at double triangle vertex Case 2.\\ Double triangle reduction transforms left to right.\\Decompletion at $\infty$ transforms top to bottom.}
  \label{fig:DTR2}
\end{figure}

\tpoint[(Case 2)]{Lemma} \label{DTRCaseTwo}
\statement[eq]{
  Let $G$ be a connected 4-regular graph and $G'$ the double triangle reduction of $G$.
  Suppose we decomplete both graphs at one of the tips of the double triangle (see Figure~\ref{fig:DTR2}), denoted $\widetilde{G}$ and $\widetilde{G'}$ respectively.
  Then
  \[ ^5\Psi_{\widetilde{G}}(1,\, 3,\, 4,\, 5,\, 2) = \pm\, ^3\Psi_{\widetilde{G'}}(4,\, 5,\, 1).\]
}

\begin{proof}
  To prove the equality, we use the same techniques as in Case 1.
  From a 5-invariant of $\widetilde{G}$, as $\{1,\, 2,\, 3\}$ forms a triangle, we have
  \[
    \pm^5\Psi_{\widetilde{G}}(1,\, 3,\, 4,\, 5,\, 2)
      = \Psi_2^{13,45}\Psi^{124,235} - \Psi^{123,245}\Psi_2^{14,35}
      = -\Psi^{123,245}\Psi_2^{14,35}.
  \]
  
  From a $D^3$ of $\widetilde{G'}$, we have (see Eq.~\eqref{D3})
  \[
  ^3\Psi_{\widetilde{G'}}(4,\, 5,\, 1)
      = \Psi^{14,15}\Psi_1^{4,5}.
  \]
  
  From here, using the graphical interpretation of Dodgsons, we get equality automatically by noting that both $^5\Psi_{\widetilde{G}}$ and $^3\Psi_{\widetilde{G'}}$ reduce to the following (up to sign):
  \begin{center}\CaseTwoBlob\end{center}
  
  To view this equality directly from Dodgson properties, we have
  \begin{align*}
  \Psi_{\widetilde{G}}^{123,245}\Psi_{\widetilde{G},2}^{14,35}&=\Psi_{\widetilde{G}\setminus2}^{12,45}\Psi_{\widetilde{G}/2}^{14,35}=\Psi_{\widetilde{G}\setminus123/45}\Psi_{\widetilde{G}\setminus3/12}^{4,5}=\Psi_{\widetilde{G'}\setminus14/5}\Psi_{\widetilde{G'}/1}^{4,5} \\
  \Psi_{\widetilde{G'}}^{14,15}\Psi_{\widetilde{G'},1}^{4,5}&=\Psi_{\widetilde{G'}\setminus14/5}\Psi_{\widetilde{G'}/1}^{4,5}
  \end{align*}
  where we've used contraction-deletion (Proposition~\ref{contdel}) and Eq.~\eqref{2valentvertex}.
\end{proof}

\tpoint[($\mathbf{c_2}$ Decompletion with Double Triangle)]{Theorem} \label{DTRInvar}
\statement[eq]{
  Let $G$ be a connected 4-regular graph and $G'$ be the double triangle reduction of $G$.
  Then, decompleting at any (same) vertex $v$,
  \[ c_2(G - v) = c_2(G' - v). \]
}

\begin{proof}
  Theorem~\ref{denomredc2} together with Theorem~\ref{DTR}, Lemma~\ref{DTRCaseOne}, and Lemma~\ref{DTRCaseTwo} gives the result.
\end{proof}

\tpoint{Remark}\label{DTRCommute}
\statement{
  Decompletion and double triangle reduction commute in the sense that you can either decomplete first and then double triangle reduce or vice versa.
} 

Theorem~\ref{DTRInvar} also gives the following Corollary which deals with a special case of the $c_2$ completion conjecture (Conjecture~\ref{c2decompconj}).  In ~\cite{specialc2} this is the $T$ case.

\tpoint{Corollary}\label{Tcompletion}
\statement[eq]{
  Let $G$ be a connected 4-regular graph and $v$ and $w$ be adjacent vertices of $G$.
  Suppose $v$ and $w$ share two common neighbours.
  Then:
  \[ c_2(G - v) = c_2(G - w) \]
}

\begin{proof}
  Since the two decompletions are symmetric, this is the equivalence of the bottom row of Figure~\ref{fig:DTR1}, proved in Lemma~\ref{DTRCaseOne}.
\end{proof}

In general, it suffices now to prove the completion conjecture for double triangle free graphs.

Finally, many of the methods used in this section to prove the double triangle reduction invariance of $c_2$ can also be extended to exploit other graphical structures in primitive divergent graphs.  As we will see in the following section, similar configurations that lead to these free factorizations of denominators will come in handy in the actual computation of $c_2$ invariants.

\section{Computation of $c_2$ invariants at 11 loops}
We analyze primitive divergent graphs in $\phi^4$ with loop order 11. We first give some background information in the context of computing $c_2$ invariants of graphs and narrow our focus to computing $c_2$ invariants of certain families of graphs. We end up investigating $c_2$ invariants of all 1731 completed primitive, irreducible and double triangle free graphs at 11 loops.

We detail our method for computing the decompleted $c_2$ invariants of these 1731 completed primitive graphs, and then go on to discuss some of the interesting patterns that emerge from analyzing the prefixes (finite initial subsequences) of $c_2$ invariants for 11 loop $\phi^4$ graphs.

\bpoint{Computational preliminaries}%
In the context of computing the $c_2$ invariant of graphs, it is very useful to assume the $c_2$ completion conjecture (Conjecture~\ref{c2decompconj}). This is because it allows us to refer to \textit{the} decompleted $c_2$ invariant that was computed for some primitive graph $G$, so we will write  $c_2^{(q)}(\widetilde{G})$ for $c_2^{(q)}(G - v)$ where $v$ is any vertex in $G$. There is much empirical evidence for this conjecture, and it is also known for special cases, see ~\cite{specialc2} or Corollary~\ref{Tcompletion}. From here on, when calculating $c_2$ invariants, we shall \textit{always} assume this conjecture and use this notation.

The periods of reducible completed primitive graphs (Definition~\ref{def:reducible}) after decompletion are not interesting as shown by Theorem~\ref{prodsplit}. In light of the following propositions we focus our study of $c_2$ invariants to decompletions of irreducible completed primitive graphs as well.

\tpoint{Proposition} (Proposition 16 from ~\cite{modular}, assuming Conjecture~\ref{c2decompconj}) \\
\statement{
    The decompleted $c_2$ invariants of reducible completed primitive graphs vanish modulo $q$.
}

\tpoint{Proposition} (Theorem 5 from ~\cite{brown2014properties})
\statement{
    Let $G$ be a graph in $\phi^4$ with at least 4 vertices.
    If $G$ is not primitive, i.e. contains a non-trivial subdivergence, then the $c_2$ invariant vanishes modulo $q$.
}

\begin{proof}
   This is Theorem 5 from ~\cite{brown2014properties}.  The restriction of at least 4 vertices comes about because the proof in ~\cite{brown2014properties} actually shows that a $D_4$, $D_5$, or $D_6$ of the graph vanishes, and this only implies the desired result if $G$ is large enough that these denominators can be used to compute the $c_2$ invariant.  If $G$ has a 2 separation then the $D_4$ is used necessitating at least 5 edges, hence at least 4 vertices for a decompleted $\phi^4$ graphs.  If $G$ does not have a 2 separation, then the graph must have at least 8 edges so any of those denominators can be used and so no further hypotheses are needed for this case.
\end{proof}

Therefore for the rest of this section we will continue to assume that completed primitive graphs are irreducible and emphasize it when needed.

Recall the double triangle reduction defined in Section~\ref{SS:DTR}. We define the \textbf{ancestor} and \textbf{family} of a completed primitive graph.

\tpoint{Definition} 
\statement{
    If $G$ is a completed primitive graph, by Proposition 2.22 in ~\cite{census} any sequence of double triangle and product reductions terminates at a unique graph $G_A$ which may have several components. The graph $G_A$ is called the \textbf{ancestor} of $G$. It is \textbf{prime} if it is connected. The family of $G_A$ is the set of completed primitive graphs which terminate at $G_A$ after all possible double triangle and product reductions.
}

Because the $c_2$ is stable under double triangle reductions (Theorem~\ref{DTRInvar}) and vanishes for products we only want to analyze prime ancestors. There are 8687 irreducible completed primitive graphs whose decompletions have 11 loops, and 1731 of them are prime ancestors.

Part of our analysis is dedicated to extending the computations of ~\cite{modular}. They primarily studied graphs in $\phi^4$ theory but also looked at graphs that are not in $\phi^4$-theory. They studied graphs up to loop order 10 by computing $c_2$ invariants. Note that they also assumed the completion conjecture. 

We recall their definition of a graph being \textbf{modular}.

\tpoint{Definition} (Definition 21 from ~\cite{modular})
\statement{
    A completed primitive graph $G$ is \textbf{modular} if there exists a normalized Hecke eigenform $f$ for a congruence subgroup of $SL_2(\mathbb{Z})$, possibly with a non-trivial Dirichlet-character with an integral Fourier expansion
    \[
        f(\tau) = \sum_{k=0}^{\infty}b_kq^k, q = e^{2\pi i \tau}, b_k \in \mathbb{Z}
    \]
    such that the decompleted $c_2$ invariant
    satisfies
    \[
        c_2(\widetilde{G})^{(p)} \equiv -b_p \mod p
    \]
    for all primes $p$.
}

For more on modular forms including definitions of Hecke eigenforms and congruence subgroups, see ~\cite{koblitz1993introduction}.  We will not need any properties of modular forms here as we will just be testing our sequences against a list of Fourier expansion coefficient sequences of modular forms. This list was previously computed by one of us in Sage ~\cite{sagemath}.

There exist graphs that are proven to be modular, see ~\cite{k3} or ~\cite{logan}. In a computational setting, in ~\cite{modular} it was assumed that if the $c_2$ invariant that was computed for some graph matched up to some modular form for a large enough number of primes, they were confident enough to say that the graph was modular. Most modular graphs in Table~\ref{modformtable} were attained this way. We shall do the same for our analysis, and we will always mention the number of primes to which we have verified the modularity of a graph.

All of the modular forms that we consider in this study are newforms, which we describe (not uniquely) in terms of weight and level.

For example, the decompleted $c_2$ invariant of the first $4$ primes of the graph $P_{8,39}$ is $(0,-1,0,0)$, which coincides with the following q-expansion of the weight 3 and level 8 newform:

\[ q - \underbrace{2q^2}_{\equiv 0 \mod 2} - \underbrace{2q^3}_{\equiv 1 \mod 3} + 4q^4 + \underbrace{0q^5}_{\equiv 0 \mod 5} + 4q^6 + \underbrace{0q^7}_{\equiv 0 \mod 7} + O(q^8). \]

Our main method of computing $c_2$ invariants for a graph $G$ is to use Theorems~\ref{denomredc2} and~\ref{cw}. Because Theorem~\ref{cw} only works for primes we restrict ourselves to the case $q=p$ prime. Note that for any 5 distinct edges, ${}^5\Psi_G(e_1,\,\dots\,,e_5) = D^5_G(e_1,\,\dots\,,e_5)$ satisfies the hypothesis of Theorem~\ref{cw}, and so do higher $n$-invariants $D^n_G(e_1,\,\dots\,,e_n)$. Thus by Theorem~\ref{cw}, one could calculate $c_2^{(p)}(G)$ by multiplying out ${}^5 \Psi(e_1,\,\dots\,,e_5)^{p-1}$ and obtaining the coefficient of $\alpha_6^{p-1}\,\dots\,\alpha_{|E(G)|}^{p-1}$ in the resulting polynomial. This naive method would work in principle, but a five invariant for a graph at 11 loops can contain hundreds of thousands of monomials. Exponentiating to high primes quickly leads to an infeasable computation. In order to make computations feasible we proceed by

\begin{enumerate}
    \item trying to find a sequence of edges and a specific decompletion such that denominator reduction goes as far as possible and by

    \item exploiting the linear homogeneity (i.e. the polynomials are homogeneous, but has degree at most one in each indeterminate) of the Dodgson polynomials to extract $\alpha_6^{p-1}\,\dots\,\alpha_{|E(G)|}^{p-1}$ in an efficient way.
\end{enumerate}

\begin{figure}[ht]
\centering
\begin{tikzpicture}[main node/.style={inner sep=0,minimum size =.12cm,circle,fill=black!80,draw},node distance = 0.75cm,scale=1]
    \begin{scope}
    \node[] (X) at 
    (-1,1){(1)};
    \node[main node] (A) at (1,0){};
    \node[main node] (B) at (2,0.8){};
    \node[main node] (C) at (2,2){};
    \node[main node] (D) at (3,0){};
    
    \node[main node] (a) at (8,0){};
    \node[main node] (b) at (6,0){};
    \node[main node] (c) at (7,0.8){};
    \node[main node] (d) at (7,2){};

    \draw (A) -- (B) node[below,pos=0.5]{3};
    \draw (B) -- (C) node[right,pos=0.5]{2};
    \draw (B) -- (D) node[below,pos=0.5]{1};
    \draw (a) -- (c) node[below,pos=0.5]{a};
    \draw (b) -- (c) node[below,pos=0.5]{b};
    \draw (d) -- (c) node[right,pos=0.5]{c};
    \end{scope}
    \begin{scope}[shift={(0,-3)}]
    \node[] (X) at 
    (-1,1){(2)};
    \node[main node] (A) at (0,0){};
    \node[main node] (B) at (1,1){};
    \node[main node] (C) at (0,2){};
    \node[main node] (D) at (3,1){};
    \node[main node] (E) at (4,2){};
    \node[main node] (F) at (4,0){};
    
    \node[main node] (a) at (8,0){};
    \node[main node] (b) at (6,0){};
    \node[main node] (c) at (7,0.8){};
    \node[main node] (d) at (7,2){};

    \draw (A) -- (B) node[below,pos=0.5]{3};
    \draw (B) -- (C) node[above,pos=0.5]{2};
    \draw (B) -- (D) node[above,pos=0.5]{1};
    \draw (D) -- (E) node[above,pos=0.5]{4};
    \draw (D) -- (F) node[below,pos=0.5]{5};
    \draw (a) -- (c) node[below,pos=0.5]{a};
    \draw (b) -- (c) node[below,pos=0.5]{b};
    \draw (d) -- (c) node[right,pos=0.5]{c};
    \end{scope}
    \begin{scope}[shift={(0,-6)}]
    \node[] (X) at 
    (-1,1){(3)};
    \node[main node] (A) at (0,0){};
    \node[main node] (B) at (1,1){};
    \node[main node] (C) at (0,2){};
    \node[main node] (D) at (3,1){};
    \node[main node] (E) at (4,2){};
    \node[main node] (F) at (4,0){};
    
    \node[main node] (a) at (5,0){};
    \node[main node] (b) at (6,1){};
    \node[main node] (c) at (5,2){};
    \node[main node] (d) at (8,1){};
    \node[main node] (e) at (9,2){};
    \node[main node] (f) at (9,0){};
    
    \draw (A) -- (B) node[below,pos=0.5]{3};
    \draw (B) -- (C) node[above,pos=0.5]{2};
    \draw (B) -- (D) node[above,pos=0.5]{1};
    \draw (D) -- (E) node[above,pos=0.5]{4};
    \draw (D) -- (F) node[below,pos=0.5]{5};
    
    \draw (a) -- (b) node[below,pos=0.5]{c};
    \draw (b) -- (c) node[above,pos=0.5]{b};
    \draw (b) -- (d) node[above,pos=0.5]{a};
    \draw (d) -- (e) node[above,pos=0.5]{d};
    \draw (d) -- (f) node[below,pos=0.5]{e};
    \end{scope}
\end{tikzpicture}
\caption{The three decompletion substructures. (1) Corresponds to decompleting a non-triangle vertex, (2) corresponds to decompleting a vertex at an isolated triangle, and (3) decompleting the vertex which two triangles meet. }
\label{fig:decompletioncases}
\end{figure}
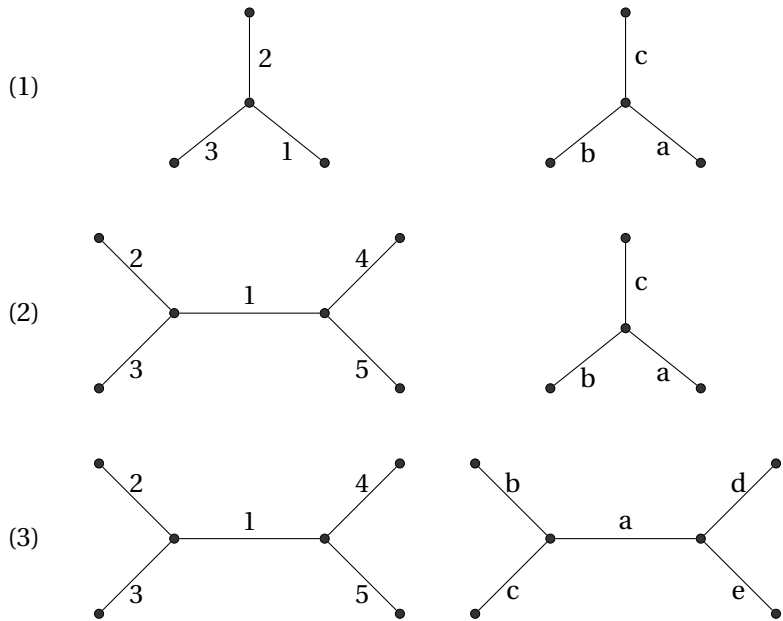

\bpoint{Simplifying graph polynomials}%
Let $G$ be a prime ancestor. We will see that it is convenient to decomplete $G$ at a vertex of a triangle, if available. Because $G$ is 4-regular and double triangle free, this gives rise to three cases of decreasing complexity: (1) $G$ has no triangle, (2) $G$ has isolated triangles, (3) $G$ has a pair of triangles that meets at one vertex. The decompletions are depicted in Figure~\ref{fig:decompletioncases}.

\tpoint{Lemma}\label{reductions}
\statement{
    Assume $G$ has a substructure as depicted in Figure~\ref{fig:decompletioncases}. Then
    \begin{enumerate}
        \item \[D^6_G(1,2,3,a,b,c) = \pm\Psi_{G/1a}^{23,bc}\Psi_{G\setminus13ac/2b}\]
        \item \[D^8_G(1,2,3,4,5,a,b,c) = \pm(\Psi_{G\setminus4/15a}^{23,bc} - \Psi_{G\setminus5/14a}^{23,bc})\Psi_{G\setminus135ac/24b}\]
        \item \[D^{10}_G(1,2,3,4,5,a,b,c,d,e) = \pm(\Psi_{G\setminus4d/15ae}^{23,bc} -\Psi_{G\setminus5d/14ae}^{23,bc} -\Psi_{G\setminus4e/15ad}^{23,bc} +\Psi_{G\setminus5e/14ad}^{23,bc})\Psi_{G\setminus135ace/24bd}.\]
    \end{enumerate}
}
Note that the right factor in all cases is the graph with all depicted edges removed. One benefit of explicit formulae in the above Lemma is that one saves a large amount of computing power otherwise needed to factorize huge polynomials.
\begin{proof}
    This is a consequence of the free factorizations, those from Proposition~\ref{dodgsonidentities} (also Proposition 3.19 from ~\cite{yeats2015some}) along with Proposition~3.25 from ~\cite{yeats2015some}.  
    Specifically, because $\{1,\,2,\,3\}$ forms a 3-valent vertex we have
  \[
    \pm^5\Psi_{G}(2,\, 3,\, c,\, b,\, 1) = \Psi_1^{23,bc}\Psi^{12c,13b}=\Psi_{G/1}^{23,bc}\Psi_{G\setminus1}^{2c,3b}.
  \]
  From Eq.~\eqref{2valentvertex} we get $\Psi_{G\setminus1}^{2c,3b}=\Psi_{G\setminus13/2}^{c,b}$.
  Because $\{a,\,b,\,c\}$ forms a 3-valent vertex we get
  \[
    D^6_{G}(1,\ 2,\, 3,\, a,\, b,\, c) = \pm\Psi_{G/1a}^{23,bc}\Psi_{G\setminus13a/2}^{c,b} = \pm\Psi_{G/1a}^{23,bc}\Psi_{G\setminus13ac/2b}.
  \]
  We prove the second case from the first. Note that $G\setminus45/1a$ has a 2-valent vertex with edges \{2,\, 3\}. In analogy to the 3-valent vertex case we have $\Psi_{G\setminus45/1a}^{23,bc}=0$.
  We have
  \begin{align*}
    \Psi_{G/1a}^{23,bc}&=\pm(\Psi_{G\setminus4/15a}^{23,bc}\alpha_4+\Psi_{G\setminus5/14a}^{23,bc}\alpha_5+X),\\
    \Psi_{G\setminus13ac/2b}&=\pm(\Psi_{G\setminus135ac/24b}(\alpha_4+\alpha_5)+Y)
  \end{align*}
  for some $X,Y$ which are constant in $\alpha_4$ and $\alpha_5$. Denominator reduction with respect to $\alpha_4$ and $\alpha_5$ gives the result.
  The third case follows from the second case in exactly the same way as the second case followed from the first.
\end{proof}

\tpoint{Remark}\label{triangle_threevalent}
\statement{For case (1) or (2) in Lemma~\ref{reductions}, we can do similar factorizations by letting $a,b,c$ be edges of a triangle instead of a 3-valent vertex. }

Of the 1731 ancestors at 11 loops, only 31 have no triangles. For these we have to stick to case (1).

We find 753 ancestors with isolated triangles and 947 ancestors with at least one pair of vertex connected triangles. Note that often after a type (2) or (3) reduction it is possible to continue the denominator reduction for some more steps.

\bpoint{Algorithm for computing $\mathbf{c_2(G)}$}%
It would be very wasteful to exponentiate these $n$-invariants outright because we only want to obtain the coefficient of $\alpha_{n+1}^{p-1}\,\dots\,\alpha_{|E(G)|}^{p-1}$. We exploit the linear homogeneity of the Dodgson polynomials for calculations. 

\tpoint[(Extraction of $\mathbf{c_2^{(p)}(G)}$ via generalized denominator reduction)]{Algorithm} \label{algoc2}

Let $G$ be the graph whose $c_2$ invariant is to be calculated. Let $\{e_1,\,\dots\,,e_n\}$ be $n$ distinct edges of the graph $G$ with the requirement the edges $\{e_1,\,\dots\,,e_n\}$ be chosen such that $D^n_G(e_1,\,\dots\,,e_n)$ is defined and can be written as a product of two polynomials which are linear in all variables.
Call these polynomials $x_1$ and $x_2$. Lemma~\ref{reductions} guarantees that this is always possible. Note that the choice of edges is not unique and that $n$ depends on how many denominator reductions the graph may allow, but $n \geq 6$. (The algorithm also works in the case of higher degrees, but it becomes very inefficient.)

\newpage\textbf{Inputs and Outputs}:
    \begin{enumerate}
        \item Input : An $n$-invariant $D^n_G(e_1,\,\dots\,,e_n)$ such that the edges $\{e_1,\,\dots\,e_n\}$ satisfy the condition described above and a sequence of edges $S$ in $E(G)\setminus \{e_1,\,\dots\,,e_n\}$, $S = (e_{n+1},e_{n+2},\,\dots\,,e_{|E(G)|})$.
        \item Output : The coefficient of $\alpha_{n+1}^{p-1}\,\dots\,\alpha_{|E(G)|}^{p-1}$ in
        $(D^n_G(e_1,\,\dots\,,e_n))^{p-1}$ modulo $p$.
    \end{enumerate}
    
    We now describe the algorithm.
    
    First we define $f_n = (D^n_G(e_1,\,\dots\,,e_n))^{p-1}$, and we get
    \begin{equation*} f_n = (D^n_G(e_1,\,\dots\,,e_n))^{p-1} = x_1^{p-1}x_2^{p-1}. \end{equation*}

    Let the edge variable $\alpha_{k}$ correspond to the edge $e_k$ for $k = n+1,\,\dots\,,|E(G)|$. We can use the linear homogeneity of each multiplicand by splitting up each term into the polynomial containing $\alpha_{n+1}$ and the polynomial not containing the $\alpha_{n+1}$. So $x_1 \mapsto y_1\alpha_{n+1} + y_2$, $x_2 \mapsto y_3\alpha_{n+1} + y_4$, and 
    \begin{equation} 
        f_n =(y_1\alpha_{n+1} + y_2)^{p-1} (y_3\alpha_{n+1} + y_4)^{p-1}. \label{almostdemred}
    \end{equation}

    We do not want to consider $f_n$ as an element of $\mathbb{Z}[\alpha_{n+1},\,\dots\,,\alpha_{|E(G)|}]$ i.e. in terms of its edge variables because the size of the polynomials $y_i$ are large and doing arithmetic would be infeasible. We consider a compactified representation. We consider each $y_i$ as a product of its irreducible factors 
    \[ y_i = \prod_j u_j \text{ is irreducible in $\mathbb{Z}[\alpha_{n+2},\,\dots\,,\alpha_{|E(G)|}$]} \]
    and $f_n$ as an element of the polynomial ring $\mathbb{Z}[u_1,\,\dots\,,u_m][\alpha_{n+1}]$ where $\{u_1,\,\dots\,,u_m\}$ are all of the irreducible factors of the multiplicands $\{y_1,y_2,y_3,y_4\}$. 
    
    Furthermore, $c_2^{(p)}(G)$ is the coefficient of $\alpha_{n+1}^{p-1}\,\dots\,\alpha_{|E(G)|}^{p-1}$ modulo $p$, so by the homomorphism of integer polynomial rings modulo $p$, we can work over $(\mathbb{Z}/p\mathbb{Z})[u_1,\,\dots\,,u_m][\alpha_{n+1}]$ which allows us to eliminate monomials and store smaller integers in memory. 

    Expand the polynomial and take the coefficient of $\alpha_{n+1}^{p-1}$ in $(\mathbb{Z}/p\mathbb{Z})[u_1,\,\dots\,,u_m][\alpha_{n+1}]$ and define 
    \begin{equation*}f_{n+1}(u_1,\,\dots\,,u_m) = \text{coeff. of $\alpha_{n+1}^{p-1}$ in $(y_1 \alpha_{n+1} + y_2)^{p-1} (y_3 \alpha_{n+1} + y_4)^{p-1}$.}\end{equation*}

    $f_{n+1}$ can be thought of as a generalized version of $(D^{n+1}_G(e_1,\,\dots\,,e_n,e_{n+1}))^{p-1}$, see Remark~\ref{grd}. Now the coefficient of $\alpha_{n+1}^{p-1}\,\dots\,\alpha_{|E(G)|}^{p-1}$ in $f_n$ is exactly the coefficient of ${\alpha_{n+2}^{p-1}}\,\dots\,\alpha_{|E(G)|}^{p-1}$ in $f_{n+1}$ modulo~$p$ when $f_n, f_{n+1}$ are considered as elements of $\mathbb{Z}[\alpha_{n+1},\,\dots\,,\alpha_{|E(G)|}]$ or $\mathbb{Z}[\alpha_{n+2},\,\dots\,,\alpha_{|E(G)|}]$, respectively. This is the end of the first iteration. 

    We proceed with the next iteration and iterate on $f_{n+1}$ instead of $f_n$. We take the variable corresponding to the next edge in the input edge sequence, $\alpha_{n+2}$. The irreducible factors of a linear homogeneous polynomial must be linear homogeneous. Therefore each $u_k$ can be expanded as 
    \begin{equation}
        \label{contdeleq}
        u_k \mapsto z_{2k-1} \alpha_{n+2} + z_{2k}.
    \end{equation}

    Substitute each
    $u_k$ in $f_{n+1}$ and let $u_{1'},\,\dots\,,u_{m'}$ be the new irreducible factors of the $z_i$ polynomials of $f_{n+1}$ in $\mathbb{Z}[\alpha_{n+3},\,\dots\,,\alpha_{|E(G)|}]$. 
        
    Similarly, we expand out $f_{n+1}$ in $(\mathbb{Z}/p\mathbb{Z})[u_{1'},\,\dots\,,u_{m'}][\alpha_{n+2}]$ and take the coefficient of $\alpha_{n+2}^{p-1}$ in $f_{n+1}$ as before to get another polynomial $f_{n+2} \in (\mathbb{Z}/p\mathbb{Z})[u_{1'},\,\dots\,,u_{m'}]$. The new polynomial $f_{n+2}$ represents the coefficient of $\alpha_{n+1}^{p-1}\alpha_{n+2}^{p-1}$ of $f_n$. Iterate this procedure until all edge variables have been eliminated, yielding the integer coefficient of $\alpha_{n+1}^{p-1}\,\dots\,\alpha_{|E(G)|}^{p-1}$ in $f_n = (D^n_G(e_1,\,\dots\,,e_n))^{p-1}$ modulo $p$. This concludes the algorithm.

\tpoint{Remark}\label{grd}
\statement{
    In Eq.~\eqref{almostdemred} in Algorithm~\ref{algoc2}, we note that 
    \[ 
        \text{coeff. of $\alpha_{n+1}^{p-1}$ in $(D^n_G(e_1,\,\dots\,,e_n))^{p-1}$} \equiv (y_1y_4 - y_2y_3)^{p-1} \mod p.
    \]
    This can be seen by taking the binomial expansion on the right side of Eq.~\eqref{almostdemred}, obtaining the $\alpha_{n+1}^{p-1}$ coefficient, and the identity
    \[ \binom{p-1}{k} \equiv (-1)^k \mod p \]
    for $
    0 \leq k \leq p-1$.
    Therefore, each step of the algorithm can be seen as a generalized denominator reduction which still works even if the multiplicands do not factor properly, hence the name of the algorithm.  
}

In practice, the effect of the input edge sequence on Algorithm~\ref{algoc2} is extremely important. We discuss this in Section~\ref{method}. We are essentially repeatedly using Proposition~\ref{contdel} in the above algorithm at Eq.~\eqref{contdeleq}. The representation of the initial Dodgson polynomials in the above algorithm as a product of their irreducible Dodgson factors $u$ muddies up the contraction-deletion interpretation of the irreducible factors in terms of Dodgson polynomials, but often these Dodgson polynomials do not factor in the earlier stages of this algorithm.

\begin{figure}[ht] 
    \centering
    \includegraphics[scale=0.125]{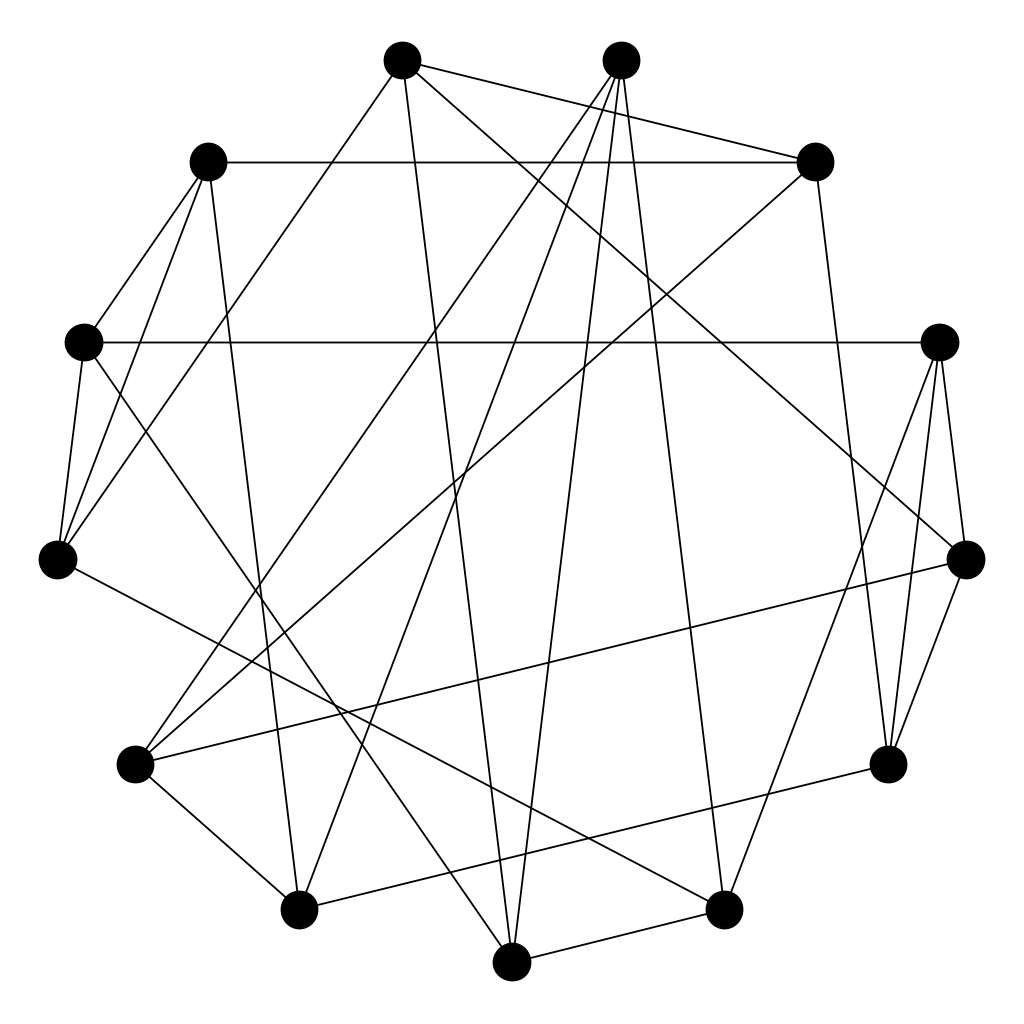}
    \[-c_2(\widetilde{P}_{11,7870}) = 1,1,3,6,1,3\]
    \includegraphics[scale=0.125]{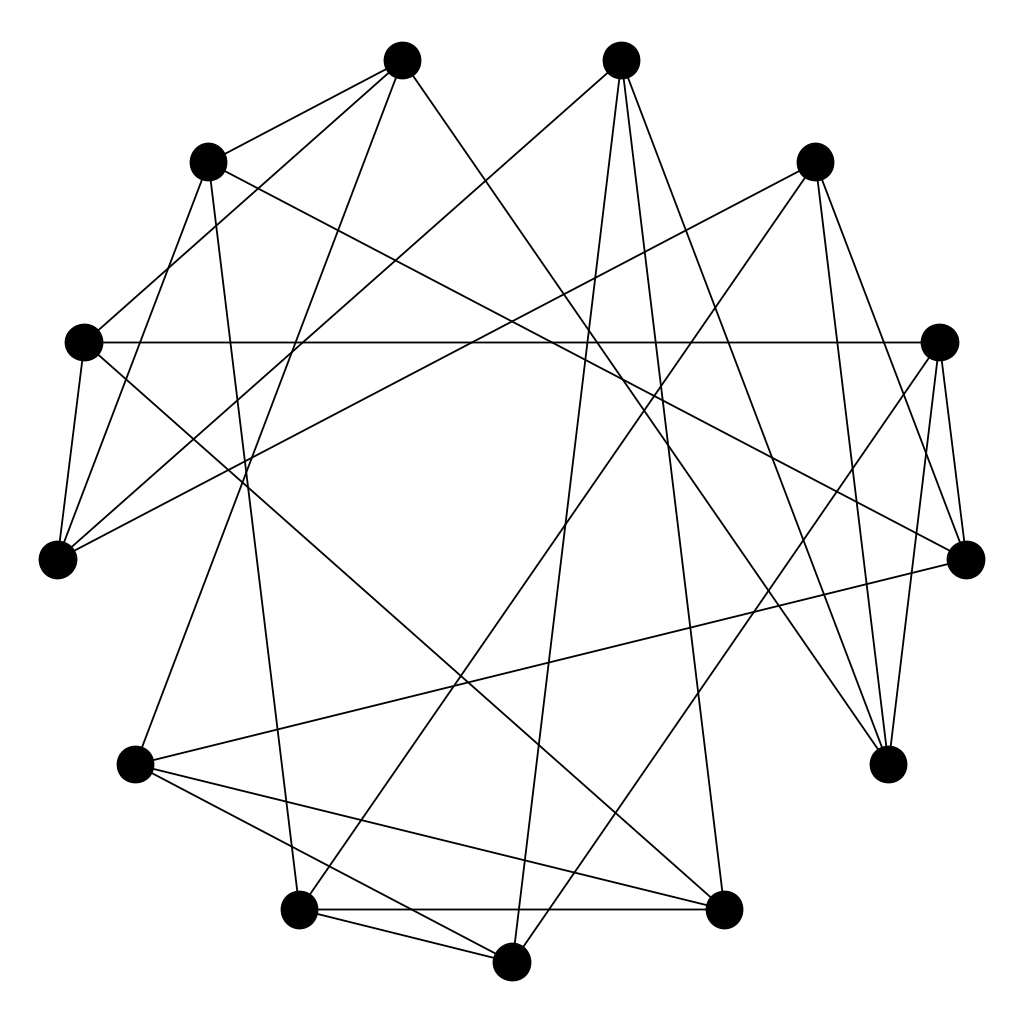}   
    \[-c_2(\widetilde{P}_{11,8684}) = 0,1,3,4,10\footnote{The prime 11 was calculated using the {\tt c2.sh} shell script of ~\cite{Hlogproc}.}\]
    \caption{The completed primitive graphs corresponding to $P_{11,7870}$ and $P_{11,8684}$. The decompleted $c_2$ for $P_{11,7870}$ took half of a second  to calculate up to $p=7$, the other one took hours for $p = 7$.} 
    \label{easyhard}
\end{figure}

Naively one would expect that the number of Dodgson factors $u$ grows like $2^k$ at step $k$ of the algorithm. The important observation is that due to the huge number of identities between sub-quotient graphs and their Dodgson polynomials this is very wrong.
Even in the hardest cases at 11 loops there exist edge-sequences so that the maximum number of Dodgson factors $u$ at every step hardly exceeds 30. In the bad case of $P_{11,8684}$ in Figure~\ref{easyhard} we used a sequence with a maximum number of 32 Dodgson factors. The polynomial in these Dodgson factors that represents the generalized denominator reduction, however, can have several million terms.

In light of Proposition~\ref{dodgsonidentities} we want to eliminate edges that cut vertices and go around triangles whenever we can. This eliminates intermediate factors and leads to a smaller intermediate polynomial in $u_1,\,\dots\,,u_m$ which speeds up computation. See Section~\ref{method} for how we generated a good sequence of edges.\footnote{After this work was finished, a procedure in ~\cite{Hlogproc} was implemented that determines the number $n_k$ of Dodgsons factors $u$ at step $k$ in the algorithm (without doing the full reduction). The procedure finds a sequence which minimizes $\max_kn_k$ within reasonable time.}

The effectiveness of this algorithm depends heavily on the graph structure. For the graph $P_{11,7870}$, shown in Figure~\ref{easyhard}, computation at $p=11$ was instant and we could compute higher primes in seconds. Note that at the top right vertex of $P_{11,7870}$
in Figure~\ref{easyhard} connects two triangles, so that we are in case (3) of Lemma~\ref{reductions}.

For the other graph in the same figure, $P_{11,8684}$, the decompleted $c_2$ invariant could not be computed at $p = 11$. Furthermore, computation at $p=7$ took 150 gigabytes of memory and about a day to compute.\footnote{After this work was finished, one of us found an improved denominator reduction algorithm which reduces a minimum of 9 edges (in contrast to 6 of case (1) in Lemma~\ref{reductions}). With this improvement the result for $p=7$ could be confirmed by point-counting on an office PC in 90 minutes ~\cite{qdr,Hlogproc}} This was one of the longest and most expensive calculations out of all of our graphs at 11 loops. Both of these calculations were done with the best possible sequence of edges we were able to find. See Figure~\ref{easyhard} for the graphs.

Finally, we also note the differences between our method using Algorithm~\ref{algoc2} and the previous method used in ~\cite{modular}. They used a point counting method, which scales exponentially with the number of edge variables left after denominator reduction. Our method instead scales exponentially with the value of $p$. If one wants to compute to high primes, their method is not limited by memory and scales better in parallel. However, our method is suitable for graphs that are harder to denominator reduce at moderate primes.

Note that possibly the main mystery of $\phi^4$ $c_2$s is the absence of certain sequences seen in non $\phi^4$ $c_2$s (see e.g.\ Table~\ref{modformtable}). The Chevalley-Warning counting is a method to produce reduced lists of possible counter-examples to even higher loop order (say 12 or 13 loops where one has 7101 or 55401 prime ancestors). A quick reduction of the number of possible counter-examples is very welcome given the high number of ancestors at high loop order. After the reduction one may use brute force counting to further reduce the lists of possible counter-examples.

\bpoint{Results}%
We were able to obtain the $c_2(G)$ for every single primitive divergent graph at 11 loops for the first 4 primes. However, we ran into memory issues at $p = 11$ for certain graphs. Computations at $p=7$ could take less than half of a second to compute for the easier graphs but around 24 hours and 150 gigabytes of RAM for the hardest graphs. Table~\ref{computedprimestable} shows how many $c_2$~invariants we calculated at the specific prime $p$. 

\begin{figure}[ht] 
    \centering
    \includegraphics[scale=0.125]{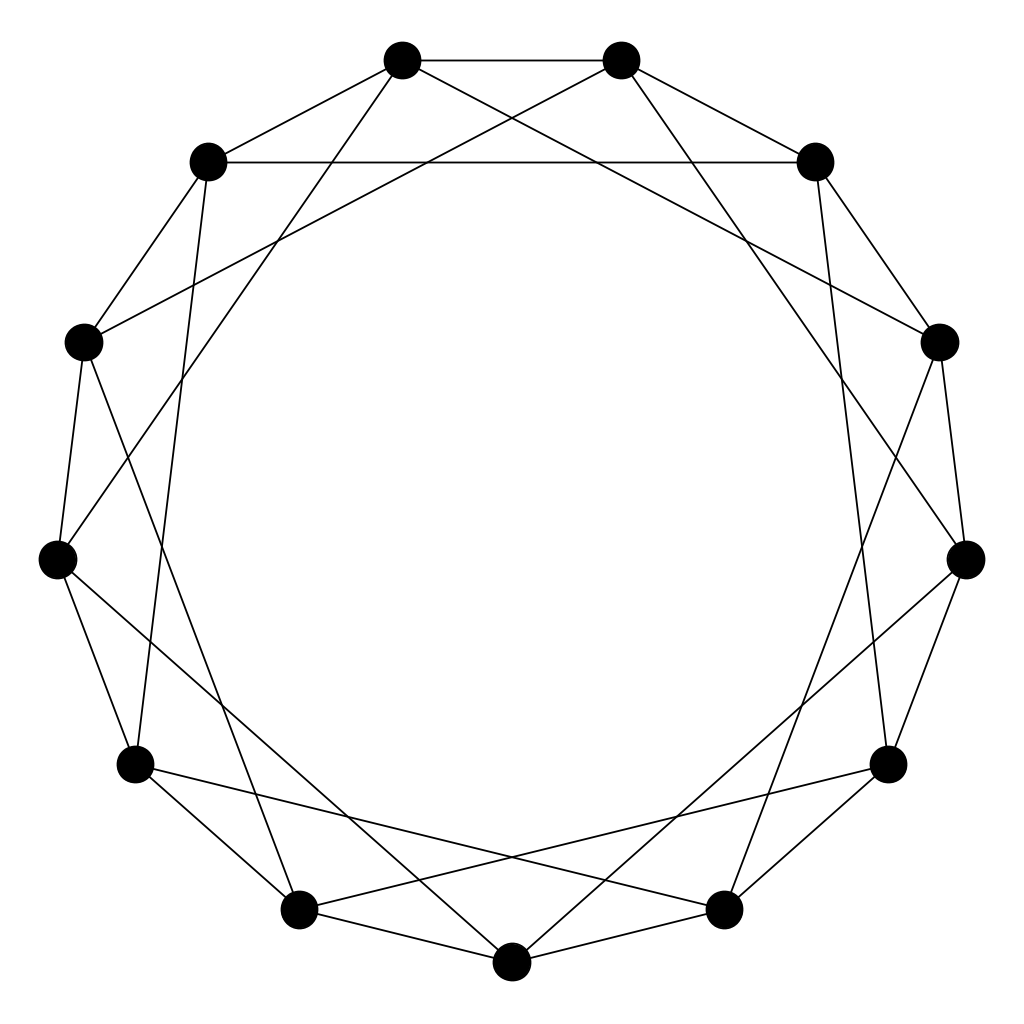}
    \[-c_2(\widetilde{P}_{11,8666}) = 1,0,1,2\] 
    \includegraphics[scale=0.125]{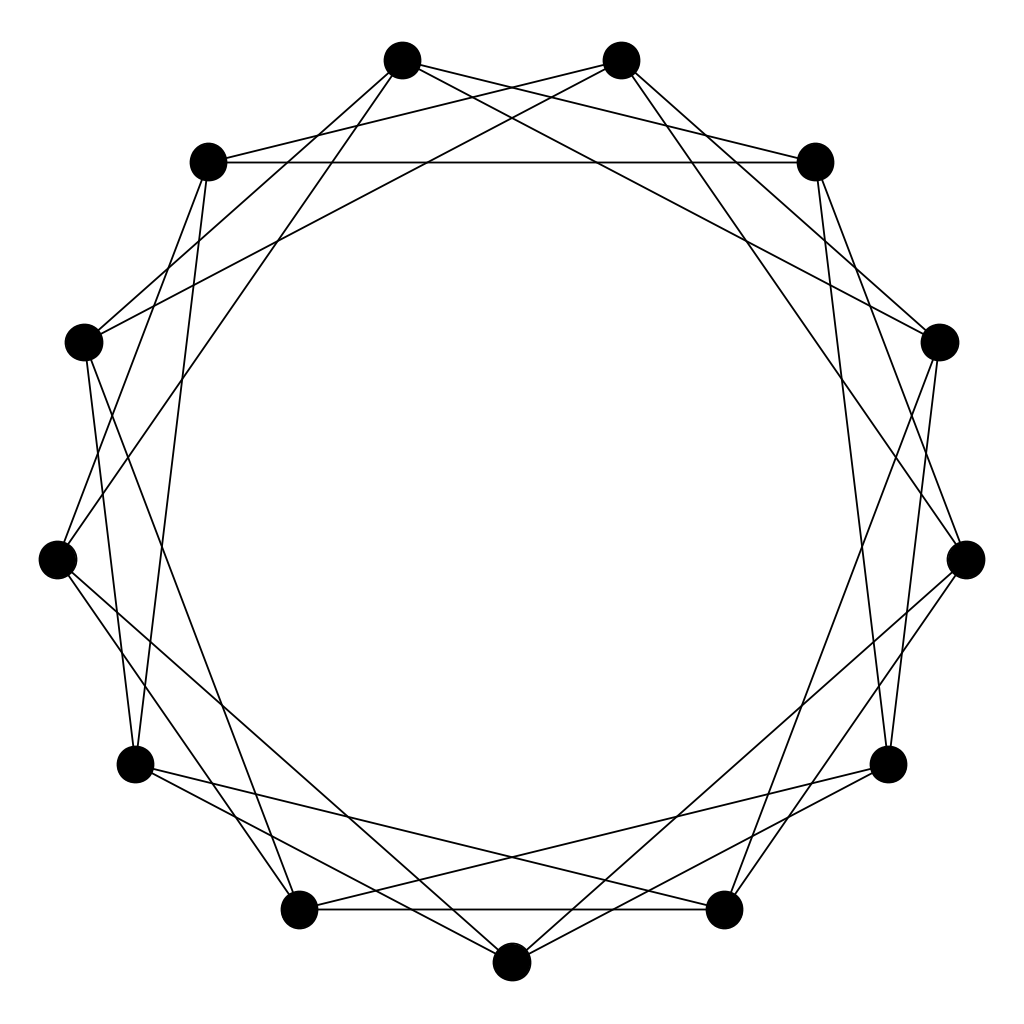}
    \[-c_2(\widetilde{P}_{11,8687}) = 0,0,3,0\]
    \caption{The circulant graphs $P_{11,8666} = C_{13}(1,3)$ and $P_{11,8687} = C_{13}(2,3)$. The decompleted $c_2$ calculations at $p=7$ took about 20 seconds for $C_{13}(1,3)$, and a few hours for $C_{13}(2,3)$.}
    \label{fig:circulants}
\end{figure}

An interesting part of these calculations is the interplay between the complexity of a graph with respect to Algorithm~\ref{algoc2} and its combinatorial structure. We noticed that many triangles in a completed primitive graph generally corresponds to an easier calculation. This is not surprising because of Proposition~\ref{dodgsonidentities} and Lemma~\ref{reductions}. However, some graphs, such as the circulant graph $C_{13}(1,3)$ (Definition 1.1 in ~\cite{yeats2016few}), have no triangles, but the symmetric structure somehow leads to a relatively fast $c_2$ calculation compared to other graphs without triangles. On the other hand, the circulant graph $C_{13}(2,3)$ $[=C_{13}(1,5)]$ was one of the hardest graphs to compute despite its apparent symmetry. The fact that $C_{13}(2,3)$ is harder to compute than $C_{13}(1,3)$ is consistent with the results in ~\cite{yeats2016few}.  The general feeling, both here and in ~\cite{yeats2016few}, is that circulants with larger gap parameters are more difficult unless they happen to be isomorphic to easier circulants.

\begin{table}    
    \centering
    \begin{tabular}{l | l l l l l l l}
        weight & 2 & 3 & 4 & 5 & 6 & 7 & 8\\
        \hline
        level & \fbox{\textbf{{11}}}$^{\phi^{>4}}$ & \fbox{\textbf{{7}}}$_P^8$ & \fbox{\textbf{{5}}}$^8$  & \fbox{\textbf{{4}}}$^9$ & \fbox{\textbf{{3}}}$^8$ & \fbox{\textbf{{3}}}$^9$ & \fbox{\textbf{{2}}}$^{10}$\\
              &  \fbox{\textbf{{14}}}$^{\phi^{>4}}$ & \fbox{\textbf{{8}}}$_P^8$ & \fbox{\textbf{{6}}}$^9$  & 7  & \fbox{\textbf{{4}}}$^9$ & 7 & 3\\
              & \fbox{\textbf{{15}}}$^{\phi^{>4}}$ & \fbox{\textbf{{11}}}$^{\phi^{>4}}$ & \fbox{\textbf{{7}}}$^{10}$  & 8 & 5 & 8 & \fbox{\textbf{{5}}}$^{10}$\\
              & 17 & \fbox{\textbf{{12}}}$_P^9$ & \fbox{\textbf{8}}$^{11}$ & 11 & 6 & 11& 6\\
              & 19 & 15 & 9 & 12 & \fbox{\textbf{{7}}}$_P^9$ & 15& 7\\
              & 20 & 15 & \fbox{\textbf{10}}$^{11}$& 15 & 8 & 15& 8\\
              & 21 & 16 & 12& 15 & 9 & 16& 8\\
              & 24 & 19 & \fbox{\textbf{{13}}}$_P^9$ & 19 & \fbox{\textbf{{10}}}$^{10}$ & 19&9\\
              & 26 & \vdots & \vdots & 20  & 10 & 20& 10\\
              & 26 & \fbox{\textbf{24}}$^{11}$ & \fbox{\textbf{{17}}}$^{10}$ & 20  & 10 & 20& 12\\
    \end{tabular}
    \caption{ The weight and level of modular graphs for 11 loops and below. All modular forms are newforms. A box indicates that a modular graph of this weight and level was found. The $\phi^{>4}$ superscript indicates that this modular form appears in non-$\phi^4$ theory, i.e. it comes from a graph with valence greater than 4. The superscript number indicates which loop order it was first found. The subscript $P$ indicates that a modular graph was found and proved to be modular for all $p$ in ~\cite{k3} or ~\cite{logan}.}
    \label{modformtable}
\end{table}

\begin{table}    
    \centering
    \renewcommand\arraystretch{2}
    \begin{tabular}{r | l l l}
        $p$ & 7 & 11 & 13\\
        \hline
        $\displaystyle\frac{\text{\# $c_2$ invariants computed at $p$}}{\text{\# prime ancestors}}$ & $\displaystyle\frac{1731}{1731}$ & $\displaystyle\frac{1424}{1731}$ & $\displaystyle\frac{751}{1731}$\\
        $\displaystyle\frac{\text{\# of distinct $c_2$ invariants}}{\text{\# of distinct sequences}}$ & $\displaystyle\frac{210}{210}$ & $\displaystyle\frac{779}{2310}$ & $\displaystyle\frac{452}{30030}$\\
    \end{tabular}
    \caption{The number of $c_2$ invariants computed up to various primes.}
    \label{computedprimestable}
\end{table}

\begin{figure}[h]
    \centering
    \includegraphics[scale=0.125]{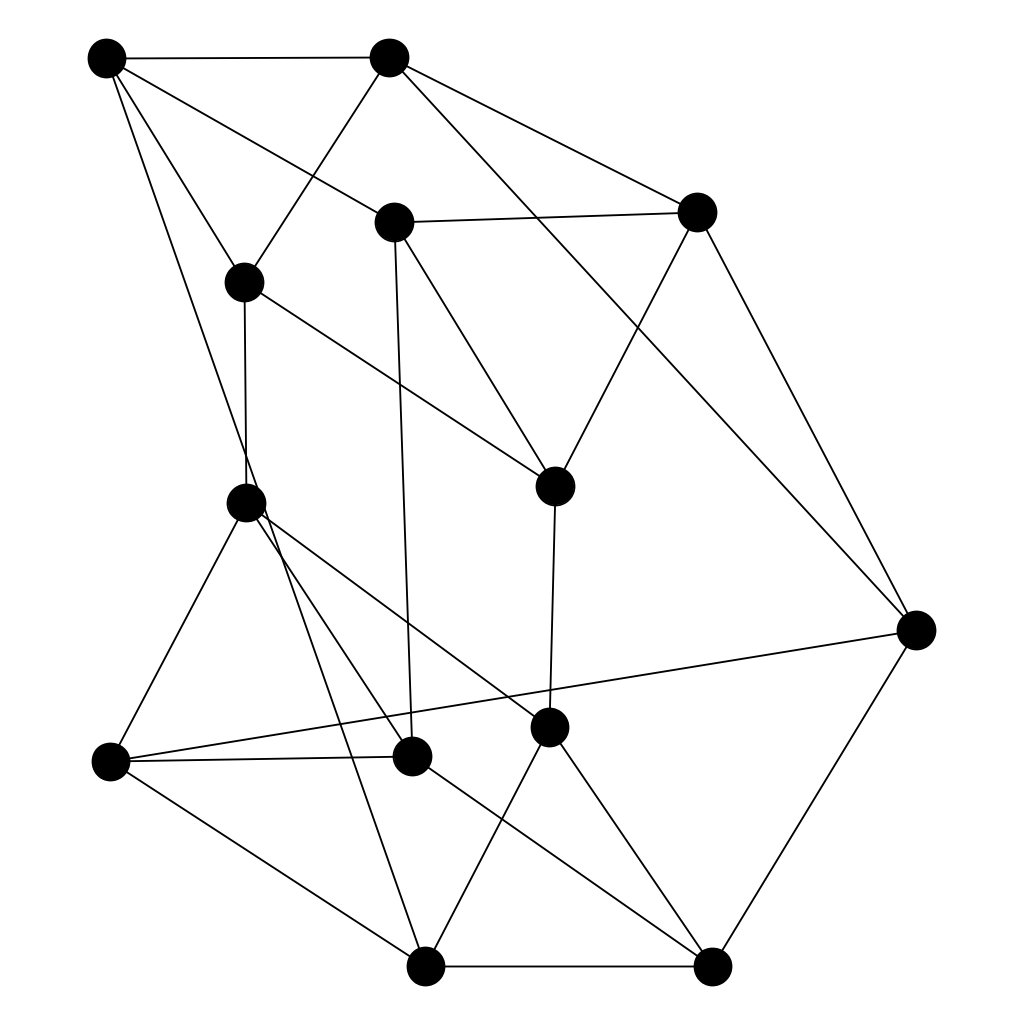} \[(4,10): -c_2(\widetilde{P}_{11,7914}) = 0,1,0,3,1,7,15,14,17,26,28 \]
    \includegraphics[scale=0.125]{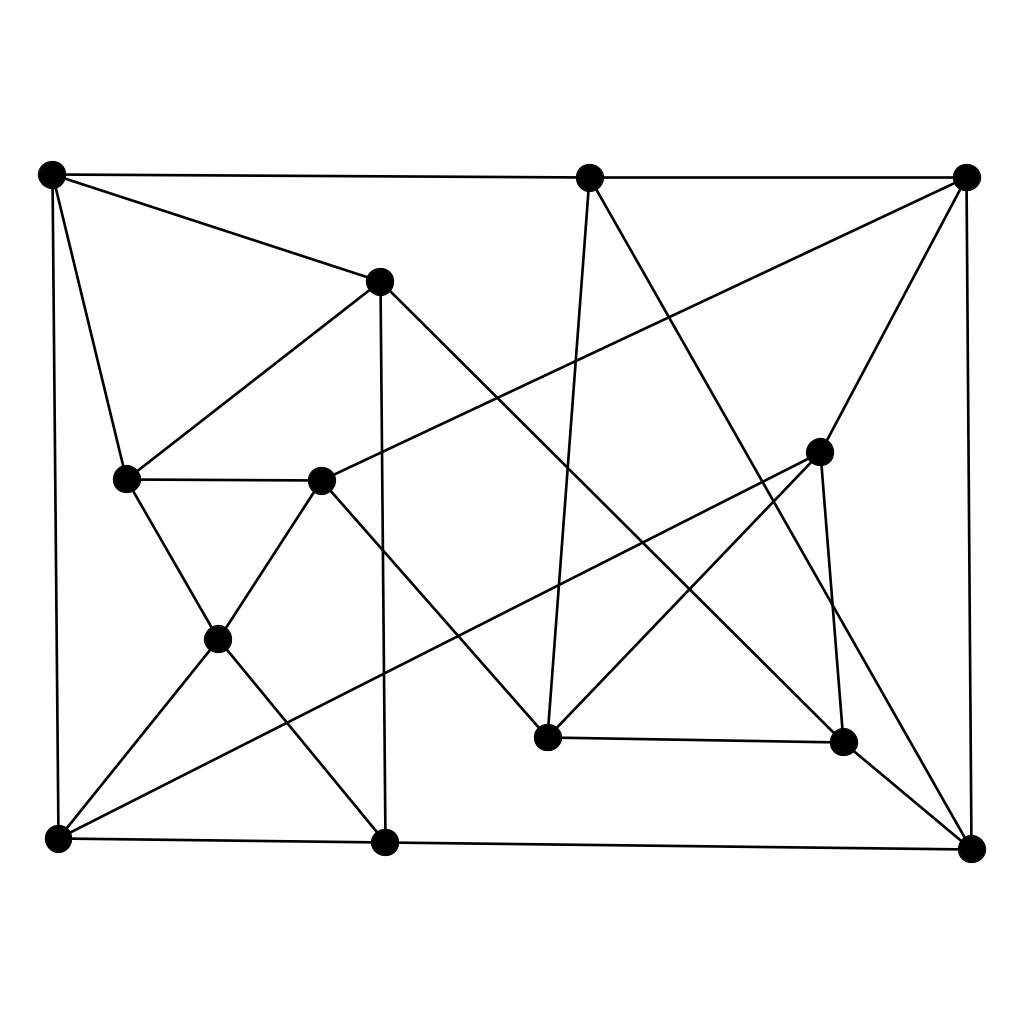} \[(3,24): -c_2(\widetilde{P}_{11,7156}) = 0,0,3,4,10,0,0,0,0,8,7\]
    \includegraphics[scale=0.125]{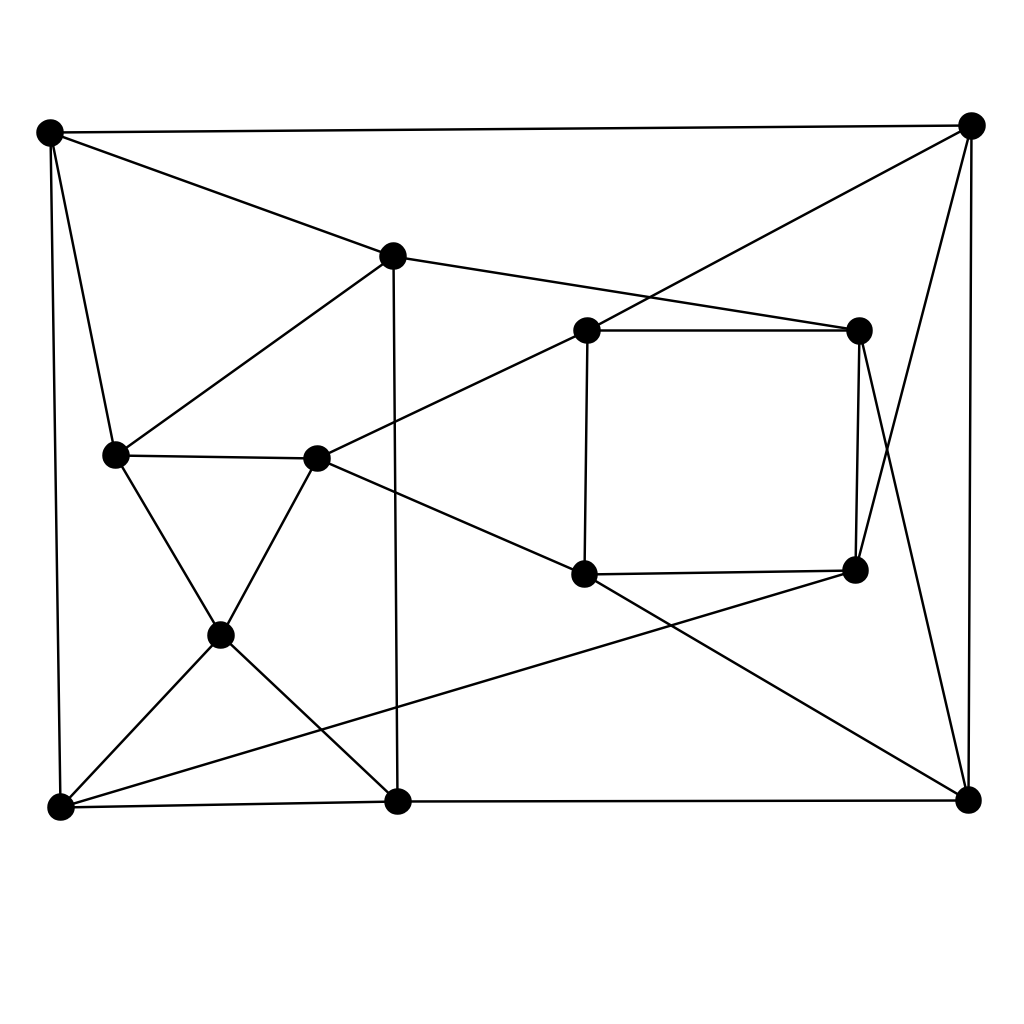}  \[(4,8): -c_2(\widetilde{P}_{11,7158}) = 0,2,3,3,0,9,16,6,13,24,26\] 
    \caption{The completed primitive graphs corresponding to the possible new modular forms.}
    \label{newmod}
\end{figure}

We found three possible new modular graphs arising from $c_2$ invariants at 11 loops. These modular graphs correspond to newforms of weight and level $(3,24),\, (4,8),$ and $(4,10)$. They are verified up to $p = 31$ for all 3 graphs. We note that $\prod_{\text{$p$ prime $\leq 31$}} p  \sim 2 \times 10^{11}$ and we are testing against 191 modular forms, so we are relatively confident that these graphs are modular. The complete description of currently known modular forms arising from $c_2$ invariants of graphs is found in Table~\ref{modformtable}. The graphs themselves can be found in Figure~\ref{newmod}. 

One particular interesting conjecture is Conjecture 26 of ~\cite{modular} which states that the modular graphs arising in $\phi^4$-theory always have weight $\geq 3$. For the $c_2$ invariants of graphs which have been calculated to $p = 13$, we find no counter-example to their conjecture up to level 46. Another interesting conjecture is part of Conjecture 25 in ~\cite{modular} which states that if $c_2(\widetilde{G}) \equiv -1 \mod p$, then the ancestor of $G$ is $K_5$. We have found no counter examples at $p=13$ for 11 loops so far for that conjecture as well.

In Table~\ref{computedprimestable} we see that for the 1424 graphs with $c_2$ up to at least $p = 11$ we get 779 distinct sequences. If we believe that every graph has a uniform probability of giving rise to any prefix of $c_2$ up to $p = 11$, letting $k = 1424$, $n = 2\times3\times5\times7\times11 = 2310$
\[
    \mathbb{E}[\text{\# of unique sequences up to $p = 11$}] = n\bigg[1 - \big[(1-\frac{1}{n})^k\big]\bigg] \sim 1063.
\]
But because of the existence of period preserving graph identities and conjectures relating $c_2$ to the period (as well as other possible $c_2$ preserving graph identities) we are not surprised at a lower number of unique sequences. However, only 62 of these 1424 graphs are related by previously known period preserving identities.

At 11 loops and just using the 1424 sequences for up to $p = 11$, we end up with at least 676 new $c_2$~invariants arising at 11 loops.

There are 145 unique $c_2$ sequences below 11 loops computed to $p = 13$. The ratio of unique sequences to prime ancestors is $\frac{145}{284} \sim 0.51$ for below 11 loops and $\frac{676 + 145}{1424 + 284} \sim 0.48$ including 11 loops.

Note that our 676 new sequences were \textbf{computed up to $p = 11$}, whereas the 145 $c_2$ sequences below 11 loops were computed \textbf{computed up to $p = 13$} in ~\cite{modular}. We expect there to be more than 676 new distinct sequences after distinguishing our sequences to $p = 13$. This would slightly change the ratios of unique sequences to prime ancestors calculated above.

We also see in Table~\ref{commseq} that certain sequences appear much more frequently than others. In particular, 14 different prime ancestors seem to share the same sequence. Of these 14 prime ancestors, only 2 share a symmetry by a known period preserving operation.
Only these two prime ancestors share the same Hepp bound. We do not know of other ways which these graphs relate to one another.

We include all of our $c_2$ invariants for 11 loop graphs in the ancillary files on the arXiv version of this paper.
They are also in the Periods file of ~\cite{Hlogproc} (which will be regularly updated).

\begin{table}[ht]    
    \centering
    \begin{tabular}{r || l|l|l|l|l|l || l}
        & \multicolumn{5}{l}{$-c_2^{(p)}(G), p = $} &     \\
        \# of occurrences & 2 & 3 &5 &7 &11 & 13 & classification \\
        \hline
        14 & 0&0&1&5&6&5& \\
        10 & 1&1&0&6&2&9& \\
        9 & 0&0&1&5&1&12& (4,6)\\
        9 & 0&2&4&3&8&3& \\
        8 & 0&0&1&0&0&9& (5,4)\\
        7 & 0&0&1&2&8&1& (6,3)\\
        7 & 0&0&2&6&10&0& \\
        7 & 0&0&0&1&3&4& (8,2)\\
        7 & 0&2&0&6&3&9& \\
        6 & 0&1&4&0&1&3& (6,7)\\
        6 & 0&1&1&2&10&9& \\
        6 & 0&1&3&2&9&1& \\
        6 & 0&1&1&1&1&1& $z_2$\\
        
        5 & 0&1&3&1&1&9& \\
        5 & 0&0&0&1&7&1&  (8,5) \\
        5 & 1&0&4&5&4&10& \\
        5 & 0&0&0&2&0&4&  (3,12) \\
        5 & 1&0&1&5&4&12& \\
        5 & 0&0&4&3&1&11& (6,4) \\
        5 & 1&2&2&0&0&4& \\
        5 & 1&0&0&1&9&7& \\
        5 & 0&2&3&3&0&9& (4,8)
    \end{tabular}
    \caption{The sequences with $\geq5$ occurrences as decompleted $c_2$ invariants of completed primitive graphs up to $p = 13$ at 11 loops. We only count sequences that have been calculated up to $p = 13$. We left the last column blank if it was not $z_n$ (see ~\cite{modular}) or a modular form.}
    \label{commseq}
\end{table}

\bpoint{Discussion}%
We discuss some of the interesting results from our investigation into the $c_2$ invariants of 11 loop graphs. We make a few heuristic observations and we also mention topics for possible further investigation.

In our results we show that the ratio of unique $c_2$ invariants to prime ancestors does not change much between different loop orders.
A related observation about the distribution of $c_2$ invariants appeared in ~\cite{yeats2018study}. One of us in ~\cite{yeats2018study} showed that the distribution of decompleted $c_2$ prefixes for the circulant graphs $C_n(1,3)$ and $C_n(2,3)$ is very uniform as we increase $n$. This uniformity, even within the same family of graphs, gave rise to the idea that maybe \textit{all} finite prefixes
show up in $\phi^4$ $c_2$ invariants if the loop order is high enough. This line of thought seems to be supported by the evidence of the present calculations.
    
As seen in Table~\ref{modformtable}, the range of modular forms is still relatively constrained at 11 loops. Based on the 751 sequences up to $p=13$, many of our modular graphs seem to arise from previously found modular forms - see Table~\ref{commseq}. Furthermore, the appearance of the two new newforms, $(4,8)$ and $(4,10)$ fits in Table~\ref{modformtable} between two newforms of the same weight. There is a possibility that some gaps in this table fill up as new $c_2$ invariants are calculated at higher loops. It seems like most of the observations made in ~\cite{modular} seem to hold for 11 loop graphs. One surprise is the newform $(3,24)$ given its relatively high level.
However, there it still seems that modular $\phi^4$ ancestors at moderate loop order have a strong preference of low levels.
    
A specific set of inputs to Algorithm~\ref{algoc2} can alter the speed of the procedure by hundreds of times - from days to minutes. We describe our heuristic method for how we chose the inputs below. It would be helpful to have more analysis done how to choose a sequence of edges that leads to a fast computation.
Proposition~\ref{dodgsonidentities}, which says to eliminate triangles and disconnect vertices, served as a starting point for our strategy of edge selection.
Note in ~\cite{Hlogproc} a simple strategy was implemented that searches for a sequence of edge-variables with the minimum number of Dodgson factors in the worst step of Algorithm~\ref{algoc2}. We did not use this strategy here.

\bpoint{Method} \label{method}%
The database of modular forms had previously been computed by one of us and the generation of it is described in Section 6 of ~\cite{modular}. 

To compute $c_2^{(p)}(G)$ for each 11-loop graph $G$, we first need to generate the inputs to Algorithm~\ref{algoc2}. The inputs are an $n-$invariant $D^n_G(e_1,\,\dots\,,e_n)$ and a sequence of edges for $E(G)\setminus\{e_1,\,\dots\,,e_n\}$. We have freedom in choosing sequences of edges and how to construct the $n-$invariant, so we want to try and select an input that leads to a fast computation. 

To construct an $n-$invariant we first use Lemma~\ref{reductions} and then denominator reduce using free factorizations from Proposition~\ref{dodgsonidentities}. For a graph with at least one triangle, we always use case (2) or case (3) in Lemma~\ref{reductions}. Otherwise, we use case (1). 

To choose an ordering of edges after constructing $D^n_G(e_1,\,\dots\,,e_n)$, we do the following: pick an edge $e_{n+1}$ that is incident on some vertex of $G \setminus \{e_1,\,\dots\,,e_n\}$ with the lowest degree, and set this as the first edge in the sequence. Proceed by picking an edge $e_{n+2}$ that is incident on some vertex of $G \setminus \{e_1,\,\dots\,,e_n,e_{n+1}\}$ with the lowest degree. Repeat this procedure until there are no edges left. Alternatively, one can choose to prioritize eliminating edges that are part of cycles.

The idea to finding a good input is to generate a set of reasonable candidate inputs, test each input by running Algorithm~\ref{algoc2} at $p=3$, and then select the best one to run at higher values of $p$. To do this, we: (1) choose different sets of edges to give to Lemma~\ref{reductions}, and (2) after using Lemma~\ref{reductions}, choose different sets of edges which are compatible with the general strategy described in the previous paragraph.

For the case of a graph with no triangles, since we are restricted to case (1) in Lemma~\ref{reductions} we can choose to decomplete at any vertex and not just at a triangle. Therefore we just obtain a set of candidate inputs by decompleting at different vertices and selecting different pairs of 3-valent vertices for case (1) in Lemma~\ref{reductions}.

Algorithm~\ref{algoc2} is very volatile with respect to a certain input.
For the modular graph $P_{11,7156}$ in Figure~\ref{newmod} the input that ended up being the best prioritized choosing edges incident on vertices with small degree.
We managed to compute $c_2(\widetilde{P}_{11,7156})$ up to $p=31$. Furthermore $p = 11$ took around 1 second this input. However if we chose another sequence of edges, $p = 11$ for the same graph did not finish within 10 minutes and we estimated that it would take at least a few hours. 

We used the symbolic library Giac ~\cite{giac} to generate the sets of $n$-invariants and edge sequences. We observed that Giac could do symbolic determinants for Dodgson polynomials much faster than other programs. We used Maple ~\cite{maple} to denominator reduce and factor polynomials. We then input the edge sequence and $D^n_G(e_1,\,\dots\,,e_n)$ to a custom C++ program created to run Algorithm~\ref{algoc2} efficiently at higher primes. 

Computation was done at the University of Waterloo on the Math Faculty Computing Facility (MFCF) specialty research servers. The specifications for the three machines we used are listed in Table~\ref{cpu_specs}. See ~\cite{waterlooservers} for more information about the MFCF servers.

\begin{table}[ht]    
    \centering
    \begin{tabular}{|l|l|l|}
    \hline
        Make/model&CPUs&Memory\\
        \hline SGI Altix XE H2106-G7&Four AMD Opteron 6168 12-core 2.3 GHz&256 GB\\
        \hline Dell PowerEdge R815&Four AMD Opteron 6276 16-core 2.3 GHz&512 GB\\
        \hline Dell PowerEdge M830&Four Intel Xeon E5-4660v3 2.1 GHz 14-core (Haswell)&256 GB\\
        \hline
    \end{tabular}
    \caption{The specifications for the three machines used for our computations.} 
    \label{cpu_specs}
\end{table}

\newpage
\section*{Appendices}\label{S:appendix}

\apoint{Table of new period identities}\label{app:FTtable}%
Using the same notation for graphs as in ~\cite{coaction} we get the following list (Table~\ref{tab:results}) of new period identities between subsets of a class of graphs based on the Fourier split. 
We used Sage ~\cite{sagemath} to implement the Fourier split transform.

Each row of Table~\ref{tab:results} corresponds to the existence of at least one Fourier split identity between a pair of graphs, one from each set. Each graph within a set (column) can currently be linked to another in that set via a series of Fourier or twist identities.

Note that all these new results are indeed proven period identities and they preserve $c_2$ invariants and Hepp bounds.

\renewcommand\arraystretch{1.1}
\begin{longtable}[c]{lccc}
  \caption{New period identities within classes up to $\ell=11$ given by Fourier split \\
  $*$ indicates that the graphs given in that row have equal periods by the Fourier split, but that these graphs form only a proper subset of a class of graphs with the same Hepp bound (see Section~\ref{SS:FTresults}), and so is an incomplete unexplained identity compared to the Hepp bound.}
  \label{tab:results} \\
  \hline
  $\ell = 9$ & $\{ P_{9,45} \}$ & $\{ P_{9,62} \}$ \\
  \hline
  $\ell = 10$ & $\{ P_{10,57} \}$ & $\{ P_{10,74} \}$ \\
  & $\{ P_{10,162},\; P_{10,172} \}$ & $\{ P_{10,197} \}$ \\
  & $\{ P_{10,218},\; P_{10,308} \}$ & $\{ P_{10,260} \}$ \\
  & $\{ P_{10,219},\; P_{10,309} \}$ & $\{ P_{10,267} \}$ \\
  & $\{ P_{10,234} \}$ & $\{  P_{10,331},\; P_{10,336} \}$ \\
  & $\{ P_{10,242} \}$ & $\{ P_{10,325},\; P_{10,326} \}$ \\
  & $\{ P_{10,250} \}$ & $\{ P_{10,321},\; P_{10,322}, P_{10,408} \}$ \\
  & $\{ P_{10,292},\; P_{10,348},\; P_{10,618} \}$ & $\{ P_{10,758},\; P_{10,762} \}$ \\
  & $\{ P_{10,300} \}$ & $\{ P_{10,332},\; P_{10,337} \}$ \\
  & $\{ P_{10,303} \}$ & $\{ P_{10,324},\; P_{10,327} \}$ \\
  & $\{ P_{10,428} \}$ & $\{ P_{10,439},\; P_{10,785} \}$ \\
  & $\{ P_{10,787} \}$ & $\{ P_{10,905} \}$ \\
  & $\{ P_{10,838} \}$ & $\{ P_{10,882} \}$ \\
  \hline
  $\ell = 11$ & $\{ P_{11,58} \}$ & $\{ P_{11,75} \}$ \\
  & $\{ P_{11,186},\; P_{11,196} \}$ & $\{ P_{11,221} \}$ \\
  & $\{ P_{11,261},\; P_{11,363} \}$ & $\{ P_{11,312} \}$ \\
  & $\{ P_{11,263},\; P_{11,370} \}$ & $\{ P_{11,322} \}$ \\
  & $\{ P_{11,278} \}$ & $\{ P_{11,388},\; P_{11,393} \}$ \\
  & $\{ P_{11,286} \}$ & $\{ P_{11,382},\; P_{11,383} \}$ \\
  & $\{ P_{11,294} \}$ & $\{ P_{11,378},\; P_{11,379},\; P_{11,471} \}$ \\
  & $\{ P_{11,355} \}$ & $\{ P_{11,389},\; P_{11,394} \}$ \\
  & $\{ P_{11,358} \}$ & $\{ P_{11,381},\; P_{11,384} \}$ \\
  & $\{ P_{11,534},\; P_{11,839},\; P_{11,859} \}$ & $\{ P_{11,551},\; P_{11,886} \}$ \\
  & $\{ P_{11,648},\; P_{11,658},\; P_{11,740},\; P_{11,760} \}$ & $\{ P_{11,683},\; P_{11,790} \}$ \\
  & $\{ P_{11,918},\; P_{11,961},\; P_{11,1109},\; P_{11,1122} \}$ & $\{ P_{11,1010},\; P_{11,1056} \}$ \\
  & $\{ P_{11,920},\; P_{11,959},\; P_{11,1116},\; P_{11,1170} \}$ & $\{ P_{11,1020},\; P_{11,1046} \}$ \\
  & $\{ P_{11,935},\; P_{11,984} \}$ & $\{ P_{11,1136},\; P_{11,1141},\; P_{11,1149},\; P_{11,1154} \}$ \\
  & $\{ P_{11,943},\; P_{11,976} \}$ & $\{ P_{11,1130},\; P_{11,1131},\; P_{11,1189},\; P_{11,1191} \}$ \\
  & $\{ P_{11,951},\; P_{11,992} \}$ & $\{ P_{11,1126},\; P_{11,1127},\; P_{11,1279},\; P_{11,1285} \}$ \\
  & $\{ P_{11,1096} \}$ & $\{ P_{11,1137},\; P_{11,1142},\; P_{11,1148},\; P_{11,1153} \}$ \\
  & $\{ P_{11,1099} \}$ & $\{ P_{11,1129},\; P_{11,1132},\; P_{11,1188},\; P_{11,1192} \}$ \\
  & $\{ P_{11,1333},\; P_{11,3183} \}$ & $\{ P_{11,1345},\; P_{11,2279},\; P_{11,3548} \}$ \\
  & $\{ P_{11,1377},\; P_{11,1931},\; P_{11,1949} \}$ & $\{ P_{11,2947} \}$ \\
  & $\{ P_{11,1380},\; P_{11,1721},\; P_{11,2064} \}$ & $\{ P_{11,1610} \}$ \\
  & $\{ P_{11,1381},\; P_{11,1720},\; P_{11,2065} \}$ & $\{ P_{11,1580} \}$ \\
  & $\{ P_{11,1390},\; P_{11,1960} \}$ & $\{ P_{11,1606},\; P_{11,1701} \}$ \\
  & $\{ P_{11,1400},\; P_{11,1964} \}$ & $\{ P_{11,1576},\; P_{11,1711} \}$ \\
  & $\{ P_{11,1526} \}$ & $\{ P_{11,1829},\; P_{11,1834} \}$ & $\{ P_{11,1976},\; P_{11,1980} \}$ \\
  & $\{ P_{11,1529} \}$ & $\{ P_{11,1819},\; P_{11,1825} \}$ & $\{ P_{11,2001},\; P_{11,2021} \}$ \\
  & $\{ P_{11,1738},\; P_{11,2112} \}$ & $\{ P_{11,4256} \}$ \\
  & $\{ P_{11,1846} \}$ & $\{ P_{11,2037},\; P_{11,2084} \}$ \\
  & $\{ P_{11,1849} \}$ & $\{ P_{11,1991},\; P_{11,1995} \}$ \\
  & $\{ P_{11,1850} \}$ & $\{ P_{11,1988},\; P_{11,1998} \}$ \\
  & $\{ P_{11,1863} \}$ & $\{ P_{11,2012},\; P_{11,2025} \}$ \\
  & $\{ P_{11,1864} \}$ & $\{ P_{11,2011},\; P_{11,2018} \}$ \\
  & $\{ P_{11,2305} \}$ & $\{ P_{11,2656} \}$ \\
  & $\{ P_{11,2306},\; P_{11,2607} \}$ & $\{ P_{11,2519} \}$ \\
  & $\{ P_{11,2383} \}$ & $\{ P_{11,2581} \}$ \\
  & $\{ P_{11,2438} \}$ & $\{ P_{11,2675} \}$ \\
  & $\{ P_{11,2451},\; P_{11,2577} \}$ & $\{ P_{11,2484} \}$ \\
  & $\{ P_{11,2590} \}$ & $\{ P_{11,2614} \}$ \\
  & $\{ P_{11,2881},\; P_{11,2887} \}$ & $\{ P_{11,2910} \}$ \\
  & $\{ P_{11,2933} \}$ & $\{ P_{11,2980},\; P_{11,2981},\; P_{11,5338} \}$ \\
  & $\{ P_{11,2940} \}$ & $\{ P_{11,2976},\; P_{11,2977},\; P_{11,3024},\; P_{11,5466} \}$ \\
  & $\{ P_{11,2965} \}$ & $\{ P_{11,2979},\; P_{11,2982},\; P_{11,5468} \}$ \\
  & $\{ P_{11,3035} \}$ & $\{ P_{11,3046} \}$ \\
  & $\{ P_{11,3069},\; P_{11,3075} \}$ & $\{ P_{11,3098} \}$ \\
  & $\{ P_{11,4253} \}$ & $\{ P_{11,4747},\; P_{11,4764} \}$ \\
  & $\{ P_{11,4381} \}$ & $\{ P_{11,4805},\; P_{11,4924} \}$ & $*$ \\
  & $\{ P_{11,5517} \}$ & $\{ P_{11,6094} \}$ \\
  & $\{ P_{11,5533} \}$ & $\{ P_{11,6101} \}$ \\
  & $\{ P_{11,5717} \}$ & $\{ P_{11,6181} \}$ \\
  & $\{ P_{11,5719} \}$ & $\{ P_{11,6278} \}$ & $*$\\
  & $\{ P_{11,6081} \}$ & $\{ P_{11,6336} \}$ & $*$\\
  & $\{ P_{11,6098} \}$ & $\{ P_{11,6327} \}$ & $*$\\
  & $\{ P_{11,6151} \}$ & $\{ P_{11,6342} \}$ \\
  \hline
\end{longtable}

\newpage
\apoint{New modular graphs}%
Below we list 3 graphs at 11 loops which correspond to new modular forms that have not been found at loop order less than 11. Interestingly, decompleting at the rightmost vertex for $P_{11,7914}$ yields a highly symmetric structure. 

\newpage
\printbibliography
\bigskip

\end{document}